**Title:** Study the effect of electrode material, its surface, and dielectric material on plasma and properties of plasma-activated water

**Authors**

Vikas Rathore[1,2*] and Sudhir Kumar Nema[1,2]

1. Atmospheric Plasma Division, Institute for Plasma Research (IPR), Gandhinagar, Gujarat 382428, India

2. Homi Bhabha National Institute, Training School Complex, Anushaktinagar, Mumbai 400094, India

***Email: vikas.rathore@ipr.res.in**

## Abstract

In the present work, the significance of various components (ground electrode material and its surface (knurling pitch size), power electrode material and type, and dielectric material) of dielectric barrier discharge plasma device (DBD-PD) on plasma and plasma-activated water (PAW) properties are studied. The characterization of plasma is performed by studying voltage-current waveform and plasma discharge power. In addition, the characterization of PAW is performed by studying the physicochemical properties (PP) and reactive oxygen-nitrogen species (RONS) concentration.

The results of plasma characterization and PAW properties reveals that introducing knurling to ground electrodes showed significant improve the physicochemical properties of PAW and RONS concentration. Moreover, the use of quartz over glass as dielectric layer provides a substantial enhancement in PAW properties. Furthermore, the use of wire as a power electrode compared to mesh and sheet also help in improving the PAW properties. Further, we observed that the ground and power electrodes made using copper enriches the RONS concentration and physicochemical properties of PAW compared to brass and stainless steel.

## 1. Introduction

The emerging applications of plasma activated water (PAW) in the field of plasma medicine, plasma agriculture, food preservation, and senitization industry, etc. provide it a lot of recognition all over the world[1-6]. These applications of PAW are possible due to the dissolution of various reactive oxygen-nitrogen species (RONS) in it[7-10]. The oxidizing species (reactive oxygen species, ROS) such as hydroxyl ($^{\cdot}OH$) radical, hydrogen peroxide ($H_2O_2$), superoxide ions ($O_2^-$), dissolved $O_3$, and peroxynitrite ions ($ONOO^-$), etc. provide PAW excellent antimicrobial efficacy[5, 6, 11, 12]. The antimicrobial efficacy of PAW towards bacteria, fungi, viruses, and pest has already been reported in published work of various researchers[2, 5, 6, 13]. In addition, selective killing of cancer cells and non-cytotoxic of PAW towards skin cells have also been explored in past literature [14].

This antimicrobial activity of PAW is widely used in the surface disinfection of a wide variety of food products including meat products, sea food, fruits, and vegetables, etc. In addition, no change in phenotypic characteristics and nutritional value was observed after PAW treatment with food products[1, 5, 6].

Along with reactive oxygen species (ROS), a high concentration of reactive nitrogen species (RNS) is also present in PAW in the form of nitrate ($NO_3^-$) and nitrite ($NO_2^-$) ions, etc[7-10]. Hence, being a rich source of nitrogen species, PAW can also be used as a fertilizer to enhance crop growth [15]. Past literature also demonstrated the use of PAW to enhance seeds germination and plant growth in a variety of crops[1, 4, 16, 17].

Hence, considering the above applications of PAW, different types of plasma devices are used to produce PAW. These devices are mainly composed of different geometries of electrode, power supplies (high-frequency AC, radiofrequency, and microwave, etc.), type of plasma discharge (glow discharge, filamentary discharge, spark discharge, and gliding arc discharge, etc.), etc[7, 10, 11, 14, 17-21]. Also, various PAW process parameters were also studied in detail to enhance the physicochemical properties of PAW and RONS concentration in it. These process parameters include plasma-water treatment/exposure time, plasma discharge power, different type of plasma forming gases (air, $N_2$, Ar, He, $N_2$ + $O_2$, Ar + $O_2$, He + $O_2$), gas flow rate, water stirring, and controlling water temperature, etc[4-10].

However, no emphasis is given to the components used to prepare plasma devices for PAW generation as per the best of the author's knowledge. Since, different materials of construction of electrode, dielectric, and type of electrode, etc. may significantly influence the properties of plasma and PAW. This literature gap created the basis of the present investigation. In the present work, a self-made dielectric barrier discharge plasma device (DBD-PD) is used to produce air plasma and the generated air plasma is exposed to water to produce plasma activated water. The DBD-PD has three major components which play a significant role in plasma production named ground electrode, power electrode, and dielectric cone. The variation in these components is investigated in the present study. The ground electrode made using hollow metal pipe and diamond knurling is introduced on the pipe surface. Since, diamond knurling created metal spikes hence increased the localized electric field which may help in the generation of excess plasma radicals and species[22-24]. Hence, may improve the PAW properties. In addition, the variation in knurling pitch size is also studied on plasma and PAW properties[22]. For comparison of dielectric material glass and quartz are chosen with the same dimensions and geometry[25, 26]. This is due to the frequent use of glass and quartz as dielectric materials in DBD discharge[22, 24-27]. The power electrode is made by wrapping the wire, mesh, or sheet over the dielectric cone surface[24, 26, 28]. Hence, variation in a different types of power electrode is also studied on plasma and PAW properties. At last, the different materials of construction used for making ground and power electrodes and their effect on PAW physicochemical properties and RONS concentration in it are investigated. The chosen materials are copper (Cu), brass (Br), and stainless steel (SS), these materials are chosen due to their frequent use in the preparation of electrodes for various plasma devices [24, 29, 30].

The produced air plasma using DBD-PD is characterized by studying the voltage-current characteristics and air plasma species are identified using optical emission spectroscopy. The effect of variation in DBD-PD components is studied by studying the voltage-discharge current characteristics, plasma discharge power, physicochemical properties of PAW, and RONS concentration in it.

## 2. Materials and Methods

### 2.1 Experimental setup

The schematic of the experimental setup is shown in figure 1. It shows the electrical and optical characterization of plasma device, and the production of plasma activated water (PAW) using plasma device. The plasma device composes a coaxial cylindrical dielectric cone (or tube) (see figure 1) with an outer diameter 24 mm and thickness 2 mm. The cone has a double-side B24 male socket. The top side of the cone is fitted in a B24 socket receiver adapter with an air leak tube which uses an air inlet in a plasma device (see figure 1). The ground electrode of the plasma device was made using a hollow metal pipe with an outer diameter 16 mm with or without diamond knurling on its surface. This ground electrode was a tight fit in the teflon cap as shown in figure 1. The power electrode is either made of flexible mesh, sheet, and wire which is wrapped around a dielectric cone as shown in figure 1. The airflow rate feed to the plasma device was controlled using an air rotameter and set at 15 l min$^{-1}$. This plasma device was powered using 0-30 kV, 0-30 kHz higher voltage high-frequency power supply. The present work uses a constant frequency of 20 kHz for all experiments.

The generated air plasma in the plasma device was characterized using electrical and optical emission measurements. A 1000x high voltage probe (Tektronix P6015A) and a 4-channel 100 MHz, 2 GS s$^{-1}$ digital storage oscilloscope (Tektronix TDS2014C) was used to measure the applied voltage across plasma device. To measure the total current (conducion and discharge) and transported charge during plasma production, a voltage drop was measured across 31 ohms non-inductive resistor and 100 nF capacitor using a 10x voltage probe (Tektronix TPP0201) in series with the ground. The air plasma emission spectrum was measured using optical fiber and a spectrometer. The emission spectrum was measured in the range of 190 nm to 925 nm using two different spectrometers (Model EPP2000-UV from StellarNet Inc. (range 190 nm to 610 nm) and UVH-1 from ASEQ instruments (range 290 nm to 925 nm)).

A 40 ml of ultrapure milli-Q (Demineralized water, DM water) was used for PAW production. To activate the water, plasma-water inteaction time varied from 1 to 5 min. The set distance between the ground electrode tip and the water surface was 30 mm. To enhance the dissolution of reactive species produced due to plasma-water interaction, the continuous stirring and water temperature were maintained at 0 °C. The stirring of the water was controlled using a magnetic stirrer and magnetic teflon bar kept in PAW. The water stirring speed was kept constant at 300 rpm throughout the experimental duration. The water temperature was maintained at 0 °C by keeping the ice-water mixture in a storage container as shown in figure 1.

## 2.2 The role of plasma device components on plasma and PAW properties

The plasma device used to produce air plasma mainly consists of three important components which play a significant role in plasma production. The ground electrode, power electrode, and dielectric material. Hence, the present investigation emphasizes the role of these plasma device components on plasma and PAW properties. The present work investigates the role of ground electrode knurling pitch size, type of power electrode, dielectric material, and material of construction of ground and power electrode.

### 2.2.1 The ground electrode knurling

The knurling of the ground electrode may increase the localized electric field for different knurling pitch sizes. That influences the generation of radicals and species in the plasma phase. Hence, the role of different knurling pitch sizes was studied on plasma and PAW properties in the present work. The different knurling pitch sizes chosen were 0 mm, 0.5 mm, 1 mm, and 2 mm, respectively (see figure 1). While studying the role of knurling pitch size other plasma device components and PAW process parameters were kept constant.

### 2.2.2 The dielectric material

The dielectric material is one of the most important parameters during dielectric barrier discharge. To study the role of dielectric material on plasma and PAW properties two different types of dielectric material were used named glass and quartz in form of a B24 double side cone (see figure 1). The other PAW process parameters and plasma device components remain unchanged while comparing the mentioned dielectric material.

### 2.2.3 The type of power electrode

To study the impact of power electrode type on plasma and PAW properties, three different types of power electrodes were used (mesh, sheet, and wire) keeping other variables constant. This mesh, sheet, and wire were wrapped around the dielectric cone to make a power electrode as shown in figure 1.

2.2.4 Material of ground and power electrode

Three different types of materials named copper (Cu), brass (Br), and stainless steel (SS) are used to study the role of ground and power electrode material on the physicochemical properties and RONS concentration of PAW. A detail of the same is shown in figure 1. In which, a picture of different ground electrode materials and power electrode materials is shown.

2.3 Measurement of physicochemical properties and RONS concentration of PAW

2.3.1 Physicochemical properties

A pH meter (Hanna Instrument, model HI98120) was used to measured the pH of PAW. An ORP meter (Hanna Instrument, model ORP-200) was used to determine the oxidizing tendency of PAW. The dissolved ions in PAW were measured using TDS (total dissolved solid) meter (HM digital, model AP1) and EC (electrical conductivity) meter (Contech, model CC-01), respectively. The least count of instruments used in the present work was given as 0.01 (pH), 1 mV (ORP), 0.1 µS $cm^{-1}$ (EC), and 1 ppm (TDS), respectively.

2.3.2 Reactive oxygen-nitrogen species (RONS)

The preliminary detection of RONS present in PAW was performed using a strip test for $NO_2^-$ ions (Macherey-Nagel, QUANTOFIX Nitrite) and $H_2O_2$ (Macherey-Nagel, QUANTOFIX Peroxide 25), and colorimetric test kit for $NO_3^-$ ions (Macherey-Nagel, VISOCOLOR Nitrate) and dissolved $O_3$ (Hanna Instrument, HI-38054 Ozone test kit) test kit.

The quantitative estimation of RONS concentrations in PAW was determined spectrophotometrically. A standard curve of nitrate ($NO_3^-$) ions, nitrite ($NO_2^-$) ions, and hydrogen peroxide ($H_2O_2$) was prepared to determine the unknown concentration of these species in PAW. The molar attenuation coefficient of $NO_3^-$ ions, $NO_2^-$ ions, and $H_2O_2$ were given as 0.0602 (mg $l^{-1})^{-1}$ (range of $NO_3^-$ ions is 0.61 to 6.10 mg $l^{-1}$), 0.0009 (µg $l^{-1})^{-1}$ (range of $NO_2^-$ ions is 67 to 536 µg $l^{-1}$), and 0.4857 (mmol $l^{-1}$)$^{-1}$ (range of $H_2O_2$ is 9.8 to 98 µmol $l^{-1}$), respectively[5, 7]. The concentration of unknown $NO_3^-$ ions present in PAW was determined spectrophotometrically at 220 nm using mentioned molar attenuation coefficient. The $NO_2^-$ ions react with a reaction mixture of sulfanilamide and N-(1-naphthyl) ethylenediamine dihydrochloride to give reddish-purple color (azo dye) which showed maximum absorbance at 540 nm in an acidic region. Hence, the unknown concentration of $NO_2^-$ ions in PAW was measured at 540 nm using mentioned molar attenuation coefficient. Similarly, $H_2O_2$ reacts with titanium ions (titanium (IV) oxysulfate) in the acidic region to form a yellow color complex (pertitanic acid, $H_2TiO_4$) which shows maximum absorbance at 407 nm[5, 7]. This property of $H_2O_2$ is used to determine the unknown $H_2O_2$ concentration in PAW using the mentioned molar attenuation coefficient. In addition, $NO_2^-$ ions present in PAW interfere in the determination of $H_2O_2$ concentration in PAW. The $NO_2^-$ ions reacted with $H_2O_2$ and suppress the $H_2O_2$ concentration in PAW beyond the detection limit. Hence, azide ions ($N_3^-$) in the form of sodium azide ($NaN_3$) were added to PAW which reacts with nitrate ions and degrades it so the interference in $H_2O_2$ determination can be prohibited[5, 7]. Appropriate dilution was performed to determine RONS concentration if the RONS concentration in PAW exceeded the detection limit.

An indigo colorimetry method was used to determine the dissolved $O_3$ concentration in PAW. In which, the rapid decolorization of indigo reagent occurs by dissolved $O_3$ in the acidic region. The indigo reagent was prepared using potassium indigo trisulfonate, sodium phosphate, phosphoric acid, and water, respectively. The volumetric method (equation (1)) [5, 7] used to determine the dissolved $O_3$ in PAW was given as:

$$\frac{mg}{l} of\ O_3 = \frac{100 \times \Delta A}{f \times b \times v} \quad (1)$$

Where, '$\Delta A$' is the absorbance difference in PAW and blank at 600 nm, '$b$' is the optical path length of the cell (1 cm), '$v$' is the volume of PAW, and '$f$' is the sensitivity factor (0.42), respectively.

## 2.4 Residue metal analysis in PAW

The plasma generation may erode the inner (ground) electrode material. The energetic particles collide with the ground electrode and result in erosion/sputtering. The erosion of material may dissolved in water in the form of metal residue in PAW. The residual metal analysis in PAW and control were performed using inductive couple plasma-mass spectroscopy (ICP-MS) (2000B ICP-MS, Perkin Elmer) in collision mode (Helium KED). The chosen ICP-MS method of testing were Environmental Protection Agency's (EPAs) 1638 and EPA 6020 B. As the ground electrodes were made of SS (iron and carbon alloy), Br (copper and zinc alloy), and Cu. The residual metal analyzed in PAW using ICP-MS were iron, copper, and zinc. The PAW and control sample were reconstituted in hydrochloric acid, and diluted (25X dilution) with DM water and preserve with nitric acid before ICP-MS analysis.

## 2.5 Data Analysis

All experiments were repeated atleast three times ($n \geq 3$). The results were expressed in plots and tables as mean ± standard deviation ($\mu \pm \sigma$). The statistically significant difference among the group $\mu \pm \sigma$ were estimated using one-way ANOVA followed by post-hoc test (Fisher's least significant difference (LSD)).

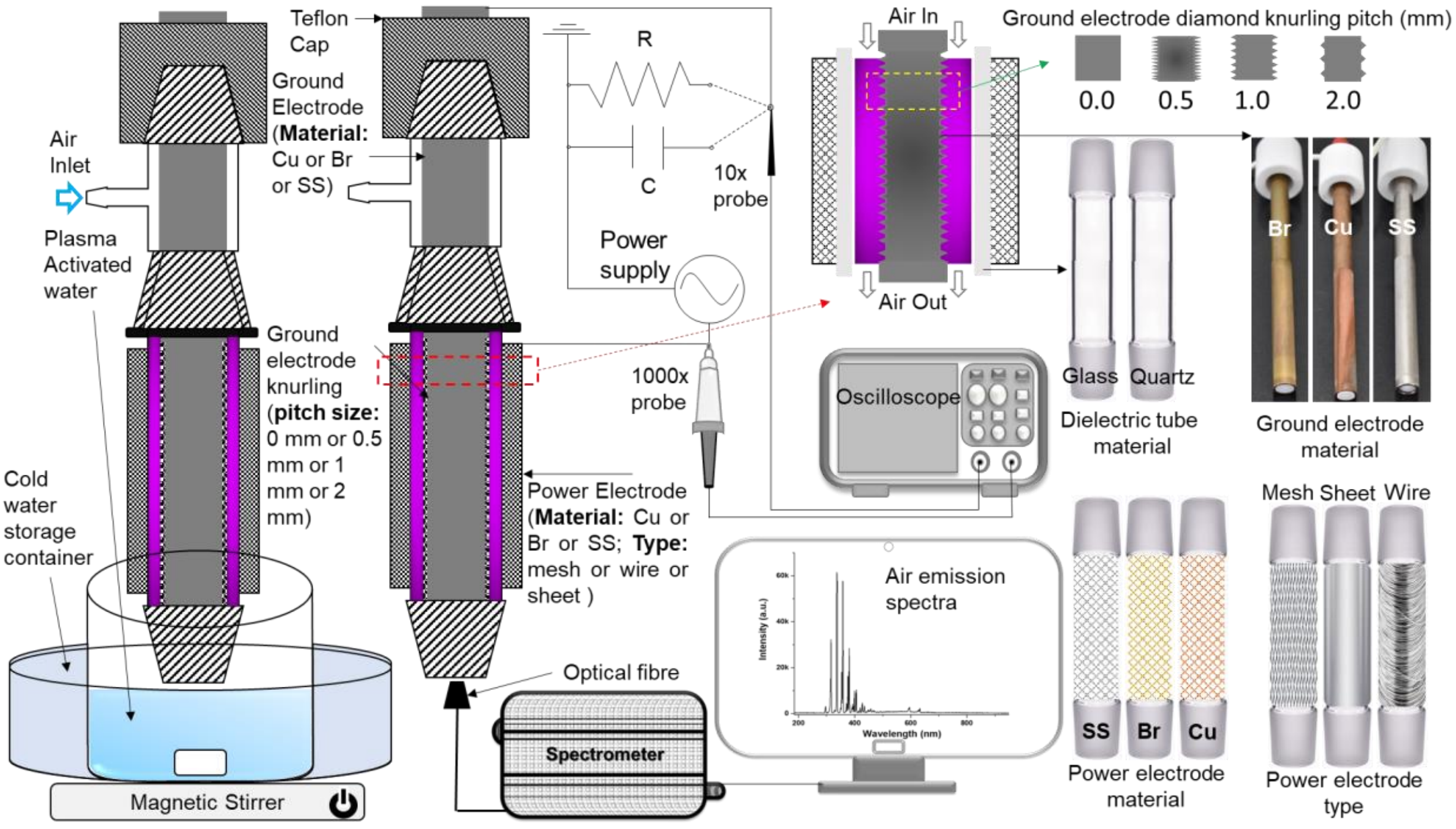

Figure 1. Schematic of electrical and optical emission characterization of plasma device and production of plasma activated water

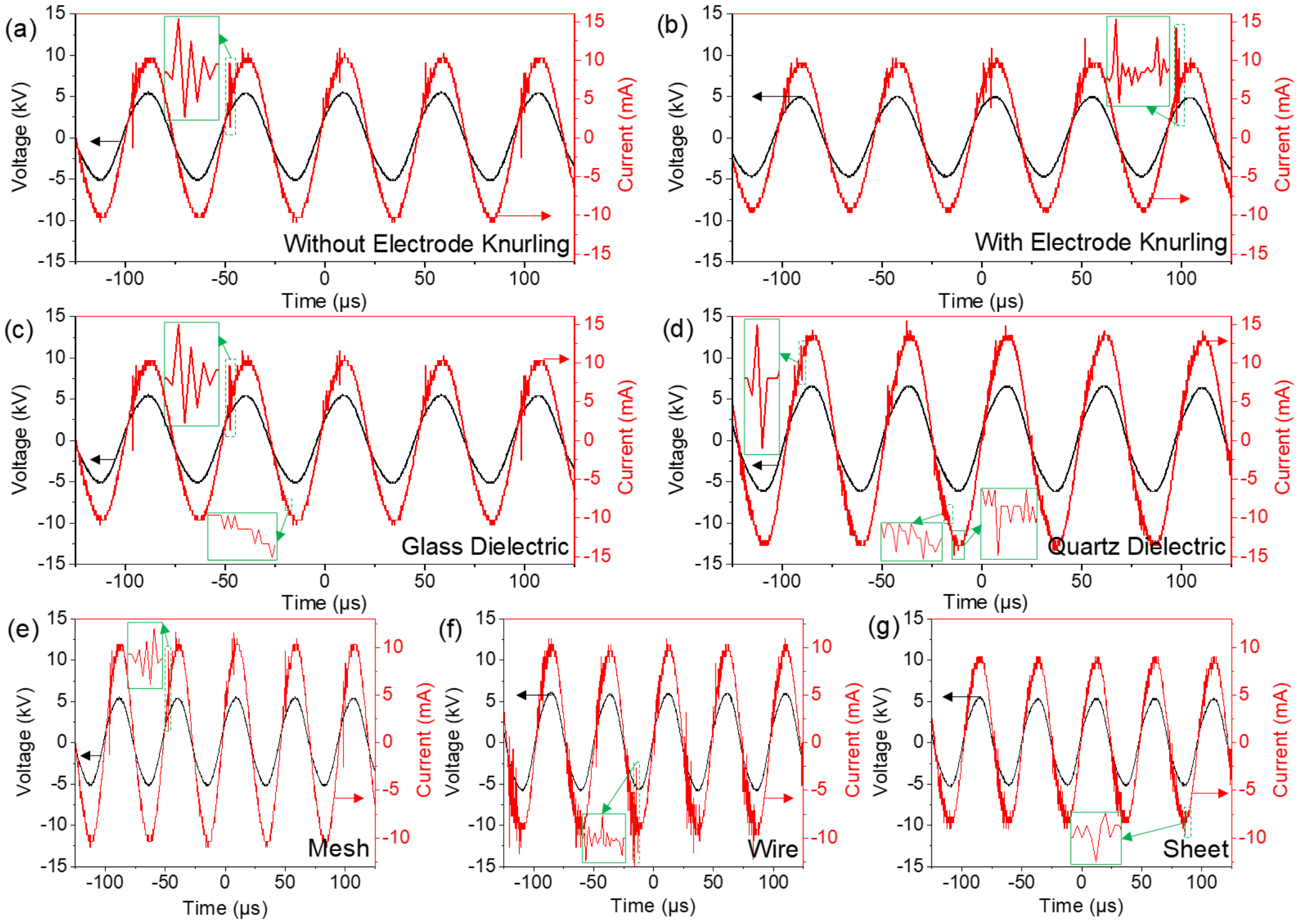

Figure 2. Voltage-current characterization of plasma device. (a, b) With and without ground electrode knurling, (c, d) variation in dielectric material (glass, quartz), (e, f, g) variation in power electrode type (mesh, wire, sheet)

## 3. Results and discussions

### 3.1 Electrical and optical emission characterization of air plasma

#### 3.1.1 Voltage-current characteristics

The variation in air plasma when produced using different electrode materials, electrode types, dielectric materials, and with and without ground electrode knurling is studied using the current waveform as shown in figure 2. The current profile shown in figure 2 is a combination of two currents, a continuous alternating current (AC) (sine wave) and a discharge current. The discharge current normally appears in each rising and falling half-cycle. The discharge current is the indicator of gas breakdown flowing through the coaxial pathway between the ground electrode and dielectric. The breakdown of gases created various ions and electrons which are indicated by various high and low multiple current peaks (filaments) in rising and falling half-cycles over continuous AC. The created ions and electrons due to gas discharge were responsible for the discharge current shown in figure 2. Hence, the periodic formation of discharge current over the continuous AC showed the formation of plasma between the ground electrode and dielectric. The time period of discharge current filaments is in order ~ 100 ns. Hence the combination of these current filaments represents the characteristics of dielectric barrier discharge (DBD) filamentary micro-discharge[31].

*Ground electrode knurling*

Figure 2 (a, b) showed the voltage-current characteristics of air plasma with and without ground electrode knurling while keeping the other design parameters and process parameters constant. The discharge current formed in the rising curve of positive half-cycle for with and without knurling ground electrode of plasma device. The peak value of discharge current with and without ground electrode knurling was 14.2 mA and 11.6 mA (positive half-cycle), respectively. The increase in discharge current peak in knurling plasma devices showed more generation of electron-ion pair during gas discharge which showed by a shoot up in the current peak. Hence, the knurling of the ground electrode supports more formation of discharge gases products. This was due to improvisation in localized electric fields with sharp knurling edges. The sharp edges of diamond knurling distorting the uniform electric field [23, 24]. Due to which localized electric field near sharp edges enhanced. As a result, generate higher pulse filaments near the sharp edges of diamond knurling. Hence, more reactive species generation occurs in the plasma phase which could be utilized for various purposes. Similar results were shown in the work reported by Mei et al.[24] and Takaki et al.[23]. However, they use screw edges and a large number of pyramids in multipoint geometry of the inner electrode to enhance localized electric field instead of diamond knurling. The diamond knurling has a slight edge over the screw-type electrode since it creates a significantly higher sharp edge density compared to the screw.

*Dielectric material*

The most commonly used dielectrics during DBD plasma production are glass and quartz. The comparison of discharge characteristics of glass and quartz dielectric, when used in plasma device is shown in figure 2 (c, d). In glass dielectric, the discharge current peaks were observed in the rising half-cycle. However, in the case of quartz dielectric, the discharge current peaks were observed in both positive and negative half-cycles. Moreover, the observed discharge current peaks in the positive rising half-cycle were significantly higher compared to the negative falling half-cycle. Hence, over a time period, two-discharge currents regions (positive rising half-cycle and negative falling half-cycle) appeared in quartz compared to the one-discharge current region (positive rising half-cycle) in the glass. This signifies a higher concentration of discharge gas products formed in plasma using quartz as a dielectric compared to glass. This will be due to better distribution of charge over quartz surface compared to glass. Since, the quartz has a crystalline structure and glass is amorphous[25]. Moreover, the use of quartz (dielectric strength 470 to 670 MV $m^{-1}$) over the glass is preferred due to its substantially higher dielectric strength compared to glass (dielectric strength 20 to 40 MV $m^{-1}$). The discharge current observation of the present investigation was also supported by work reported by Ozkan et al.[25]. In which, higher discharge current peaks were observed in quartz dielectric compared to glass dielectric. Also, filamentary discharge peaks were observed in both positive and negative half-cycles for quartz dielectric. However, for glass dielectric, observed filamentary discharge peaks in one-half cycle were substantially higher than other half cycles.

*Power electrode type*

The effect of different types of power electrodes on air plasma discharge characteristics is shown in figure 2 (e-g). The power electrode is made from mesh or winding of wire over a dielectric cone or thin metal sheet. The use of different types of power electrodes had a significant impact on discharge current characteristics. The number of high discharge-current filaments in power electrode made using wire was significantly greater compared to power electrodes made using mesh or sheet as shown in figure 2 (e-g). The discharge current filaments in the power electrode made using wire or sheet mainly occur in the negative falling half-cycle (figure 2 (f, g)). However, inverse behavior was observed in the power electrode made using mesh in which discharge current filaments appeared in the positive rising half-cycle (figure 2 (e)).

As the large number of high peaks current filaments observed in the discharge-current profile of air plasma produced using wire as power electrode showed more formation of electron-ion pairs. As a result, the density of reactive plasma species produced using wire as a power electrode will be higher compared to power electrodes made using mesh or sheet. The use of wire, mesh, and sheet (foil) as outer electrodes for DBD discharge wrapped around dielectric was previously explored by Mei et al.[24], Chang et al.[28], and Nur et al.[26]. In which, Mei et al.[24] used aluminium foil as outer electrode wrapped around dielectric have higher filamentary discharge current peaks compared to mesh. The present work also showed denser current peaks in the sheet as outer electrode compared to mesh. This was due to uniform covering of dielectric using sheet compared to mesh results in more charge accumulation on the dielectric surface and more discharge area. Hence, the number of discharge filaments increases in the sheet compared to mesh shown as dense filamentary discharge in voltage-current characteristics of the sheet as power electrode (figure 2 (g)).

In conclusion, the use of a knurled ground electrode, quartz as dielectric material, and wire as a power electrode results in the generation of a high concentration of charged plasma species/radicals. Hence, this configuration could be used in future technology where high plasma species density is required.

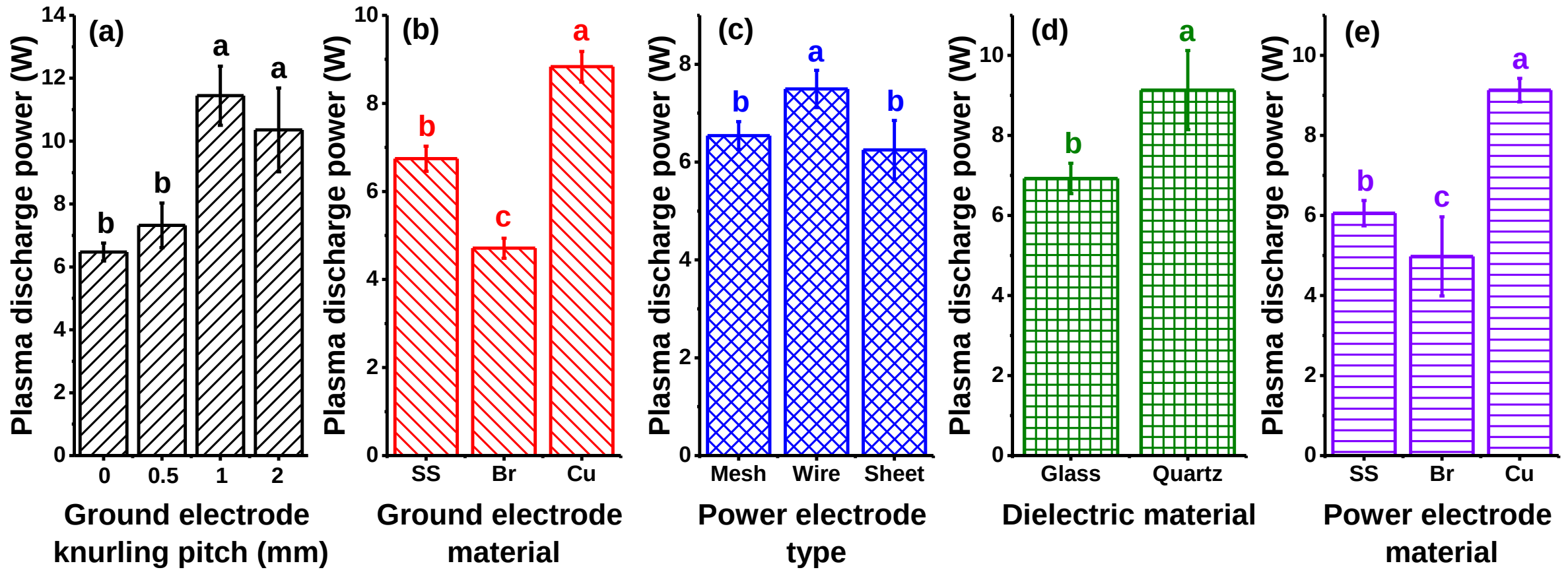

Figure 3. Variation in plasma discharge power while varying, (a) ground electrode knurling pitch, (b) ground electrode material, (c) power electrode type, (d) dielectric material, (e) power

electrode material. Different lowercase letters showed statistically significant difference ($p<0.05$, $n \geq 3$) among the group mean ± standard deviation ($\mu \pm \sigma$)

3.1.2 Plasma discharge power

As discussed above the discharge of gas is shown by the increase in the discharge-current. This discharge current signifies the movement of newly generated ionized species in the plasma. Hence, the energy is consumed during the generation of these reactive species/radicals in plasma. The higher dissipation of energy (or power) results in the formation of a high concentration of reactive species/radicals in plasma. Figure 3 (a-e) showed the variation in the plasma discharge power while varying knurling pitch, ground and power electrode material, dielectric material, and type of power electrode, respectively at constant applied voltage.

Figure 3 (a) showed the variation in the plasma discharge power with increasing knurling pitch size. An increase in knurling pitch size from 0 to 1 mm increased the plasma discharge power by 76.8%. Moreover, a further increase in the knurling pitch size from 1 mm to 2 mm decreased the plasma discharge power by 9.4%. However, this decrease in power was not statistically significant ($p > 0.05$). Hence, the knurling pitch is an important parameter that influences the generation of more reactive species/radicals in plasma. Mei et al.[24] and Takaki et al.[23] also showed improvement in energy efficiency and input energy at the same applied voltage. They used a screw-type (similar to knurling) inner electrode compared to the rode-type inner electrode and multipoint (pyramid shape) geometry (similar to knurling) compared to plane plate geometry.

The effect of ground and power electrode material on plasma discharge power is shown in figure 3 (b, e). The calculated plasma discharge power for copper (Cu) material as ground and power electrodes was substantially ($p < 0.05$) higher than brass (Br) and stainless steel (SS) ground and power electrodes. For the ground electrode, plasma discharge power for Cu was 23.7% higher compared to SS and 87.5% higher compared to Br. Similarly, for the power electrode, plasma discharge power for Cu was 50.9% higher compared to SS and 83.7% higher compared to Br. Hence, from the above discussion the choice of material of electrode while producing plasma also plays a significant role in the concentration of generated plasma species and radicals. Since higher plasma discharge signifies more formation of reactive plasma species and radicals. The role of electrode material on plasma discharge power also was reported by Jahanmiri et al.[32]. In which they showed average power consumption by Cu and Br substantially higher than SS. In addition, no significant difference was observed in power consumption when Cu and Br were used as material.

The impact of different types of power electrodes on plasma discharge power is shown in figure 3 (c). As shown in the results of discharge current characteristics, the multiple high discharge current peaks observed in power electrode made using wire compared to mesh and sheet as power electrode (figure 2 (e-f)). Similar results were observed in the plasma discharge power, in which the power electrode made using wire had 14.5% higher discharge power compared to mesh and 19.8% higher discharge power compared to the sheet. Mei et al.[24] used SS mesh and aluminium foil as outer electrode wrapped around quartz dielectric tube. They showed better energy efficacy in aluminium foil compared to SS mesh. However, in the present work, we did not observe any significant difference in the plasma discharge power when using SS sheet and SS mesh as power electrodes wrapped around the dielectric.

The effect of dielectric material on the plasma discharge power is shown in figure 3 (d). In which, the calculated plasma discharge power of air plasma when produced using quartz as dielectric significantly ($p < 0.05$) higher compared to glass as dielectric. The higher discharge power in quartz compared to glass was due to the power dissipation in each rising and falling half-cycle in quartz compared to single rising half-cycle power dissipation in the glass over a period. Similar results were observed in the work reported by Ozkan et al.[25], in which higher discharge power was observed in quartz dielectric compared to glass dielectric at constant parameters.

3.1.3 Emission spectra of air plasma

The emission spectrum of air plasma is shown in figure 4. The wavelength range shown in figure 4 lies between 185 nm to 950 nm. The recording emission spectra of air plasma mainly contain strong emission band peaks of the $N_2$ second positive system ($C\ ^3\Pi_u \rightarrow B\ ^3\Pi_g$) lies in the range of 290 nm and 440 nm (shown in figure 4 box). In addition, weak intensity $N_2^+$ ($B\ ^2\Sigma_u^+ \rightarrow X\ ^2\Sigma_g^+$) first negative system band peaks were also observed in air plasma. The observed band peaks details are shown in Table 1.

*Mechanism of formation of $N_2$ second positive system and $N_2^+$ first negative system in air plasma*

The ground state $N_2\ (X\ ^1\Sigma_g^+)_\upsilon$ present in the air was excited by direct impact excitation to upper-level $N_2\ (C\ ^3\Pi_u)_{\upsilon'}$ (equation (2)) in an applied electric field between space change developed over the dielectric surface and ground electrode. The population of upper-level $N_2$ is the result of high energy electron collision with the $N_2$ ground state[33, 34].

$$N_2\ (X\ ^1\Sigma_g^+)_\upsilon \quad + \quad e \quad \rightarrow \quad N_2\ (C\ ^3\Pi_u)_{\upsilon'} \quad + \quad e \qquad (2)$$

The radiative decay of upper-level $N_2\ (C\ ^3\Pi_u)$ to lower-level $N_2\ (B\ ^3\Pi_g)$ (equation (3)) results in the formation of strong emission band peaks of the $N_2$ second positive system as shown in Table 1.

$$N2\ (C\ ^3\Pi_u)_{\upsilon'} \quad \rightarrow \quad N_2\ (B\ ^3\Pi_g)_{\upsilon''} + h\upsilon \qquad (3)$$

The formation of excited-state $N_2^+\ (B\ ^2\Sigma_u^+)_{\upsilon'}$ from $N_2\ (X\ ^1\Sigma_g^+)_\upsilon$ ground state and its population may followed the following paths.

One-step process

$$N_2\,(X\,^1\Sigma_g^+)_\upsilon \quad + \quad e \quad \rightarrow \quad N_2^+\,(B\,^2\Sigma_u^+)_{\upsilon'} \quad + \quad 2e \qquad (4)$$

Two-step process

$$N_2\,(X\,^1\Sigma_g^+)_\upsilon \quad + \quad e \quad \rightarrow \quad N_2^+\,(X\,^2\Sigma_u^+)_{\upsilon'} \quad + \quad 2e \qquad (5)$$

$$N_2^+\,(X\,^2\Sigma_u^+)_\upsilon \quad + \quad e \quad \rightarrow \quad N_2^+\,(B\,^2\Sigma_u^+)_{\upsilon'} \quad + \quad e \qquad (6)$$

In a one-step process, electron impact ionization of $N_2\,(X\,^1\Sigma_g^+)_\upsilon$ ground state molecule occurs to $N_2^+\,(B\,^2\Sigma_u^+)_{\upsilon'}$ excited state (equation (4)). However, in a two-step process, electron impact ionization of $N_2\,(X\,^1\Sigma_g^+)_\upsilon$ ground state molecule occurs to $N_2^+\,(X\,^2\Sigma_u^+)_{\upsilon'}$ ground state (equation (5)). Then this generated $N_2^+\,(X\,^2\Sigma_u^+)_\upsilon$ ground state populated to $N_2^+\,(B\,^2\Sigma_u^+)_{\upsilon'}$ excited state with electron impact excitation (equation (6)).

The radiative decay of excited state $N_2^+\,(B\,^2\Sigma_u^+)_{\upsilon'}$ to ground state $N_2^+\,(X\,^2\Sigma_u^+)_\upsilon$ (equation (7)) results in formation of weak intensity $N_2^+\,(B\,^2\Sigma_u^+ \rightarrow X\,^2\Sigma_g^+)$ first negative system band peaks[33].

$$N_2^+\,(B\,^2\Sigma_u^+)_{\upsilon'} \quad \rightarrow \quad N_2^+\,(X\,^2\Sigma_u^+)_{\upsilon''} \quad + \quad h\upsilon \qquad (7)$$

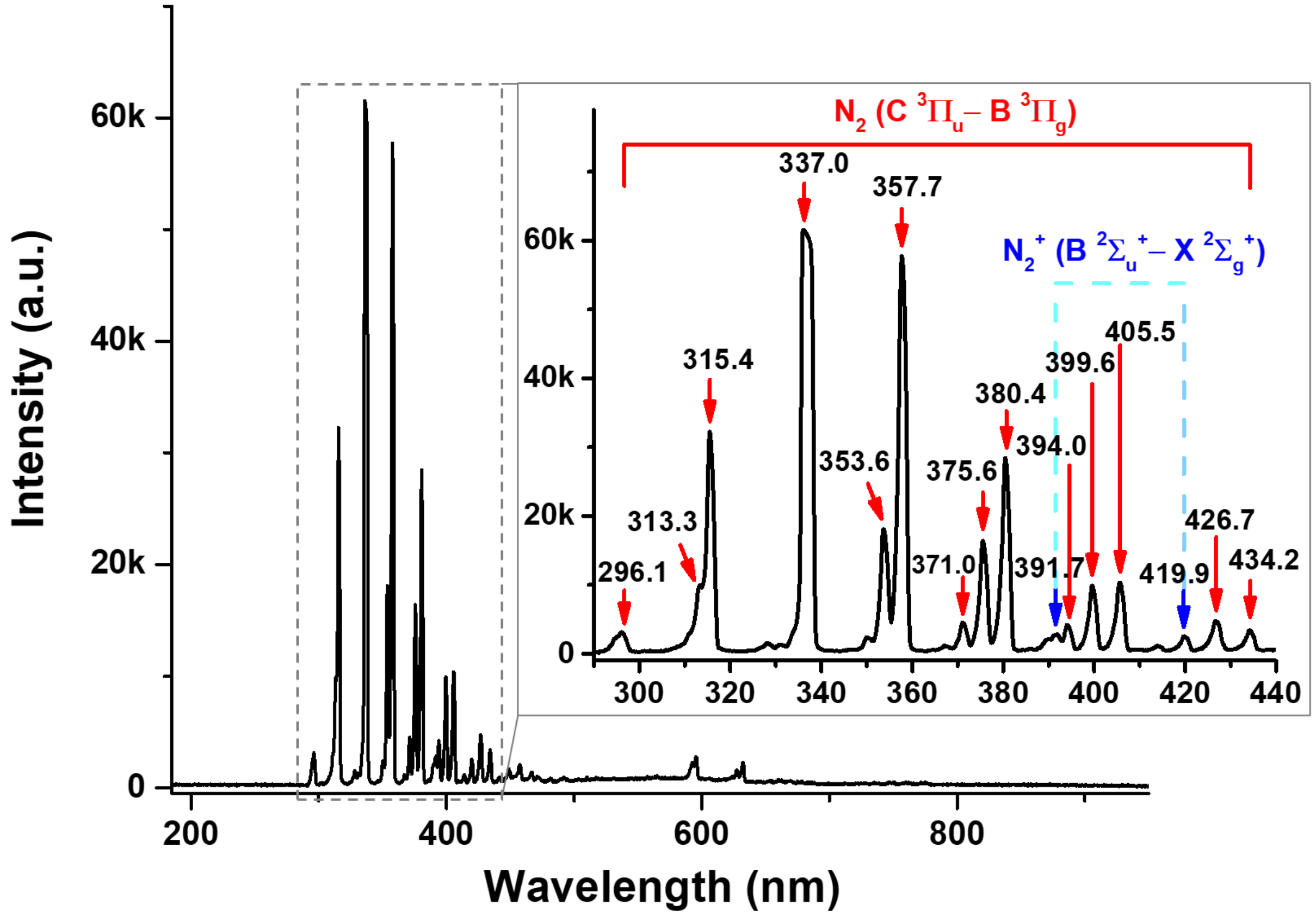

Figure 4. Optical emission spectra of air plasma

Table 1 The observed band peaks lines in air emission spectra[34]

| Species | Transitions | Spectral lines (nm) | υ' → υ" | Transition Probabilities ($s^{-1}$) |
|---|---|---|---|---|
| $N_2$ | $C\ ^3\Pi_u \rightarrow B\ ^3\Pi_g$ | 296.1 | 3 → 1 | $6.61 \times 10^6$ |
| | | 313.3 | 2 → 1 | $8.84 \times 10^6$ |
| | | 315.4 | 1 → 0 | $1.02 \times 10^7$ |
| | | 337.0 | 0 → 0 | $1.10 \times 10^7$ |
| | | 353.6 | 1 → 2 | $4.61 \times 10^6$ |
| | | 357.7 | 0 → 1 | $7.33 \times 10^6$ |
| | | 371.0 | 2 → 4 | $3.37 \times 10^6$ |
| | | 375.6 | 1 → 3 | $4.10 \times 10^6$ |
| | | 380.4 | 0 → 2 | $2.94 \times 10^6$ |
| | | 394.0 | 2 → 5 | $2.63 \times 10^6$ |
| | | 399.6 | 1 → 4 | $2.49 \times 10^6$ |
| | | 405.5 | 0 → 3 | $9.23 \times 10^5$ |
| | | 426.7 | 1 → 5 | $7.69 \times 10^5$ |
| | | 434.2 | 0 → 4 | $2.47 \times 10^5$ |
| $N_2^+$ | B $^2\Sigma_u^+$ → X $^2\Sigma_g^+$ | 391.7 | 0 → 0 | $1.10 \times 10^7$ |
| | | 419.9 | 2 → 3 | $3.13 \times 10^6$ |

3.2 Formation of RONS in water and change in physicochemical properties of water

The mechanism of formation of various reactivity oxygen-nitrogen species (RONS) in PAW is shown in Table 2. The reactions are divided into three phases – plasma phase (equations (8-20)), plasma-liquid interphase, and liquid phase (equations (21-37))[10-12, 21, 35-41]. In the plasma phase, the dissociation of $N_2$, $N_2^+$, $O_2$, and $H_2O$, etc. occurs by electrode impact dissociation into corresponding atoms (N, O, H, etc.) and molecules (OH, etc.) (equations (8-10, 12). The dissociation of molecules also occurs by high-energy excited molecules. As shown in equation (11), dissociation of $H_2O$ molecule with high energy $N_2$ molecule. Along with the dissociation reaction, the dissociative replacement also occurs by high-energy atoms. The formation of NO molecule occurs by dissociative replacement of $N_2$, $O_2$, and OH by O and N atoms (equations (13-15)). Similarly, the formation of $HO_2$ and $NO_2$ occurs by dissociative replacement of OH and NO by gases $O_3$ (equations (19, 20)). At last, recombination reactions occur in the plasma phase which results in the formation of gases $O_3$, $NO_2$, and $H_2O_2$, etc (equations (16-18)).

The region between the plasma phase and liquid phase is known as the plasma-liquid interphase. In which, the relatively long-lived species exist before dissolved into the liquid phase. These species were given as excited $N_2$, $H_2O$, $O_3$, NO, OH, $NO_2$, and $HO_2$ molecules, etc. as shown in Table 2 of plasma-liquid interphase.

The formation of stable reactive RONS is shown in the liquid phase of Table 2. The stable species which are identified and whose concentrations are measured in the present work are shown in bold (marked in red) in Table 2. The reactions (equations (21-37)) which result in the formation/degradation of stable species such as $NO_2^-$ ions, $NO_3^-$ ions, dissolved $O_3$, and $H_2O_2$ are shown in the liquid phase of Table 2. The reactions that result in the formation of $NO_2^-$ ions (equations (21,30,31,)) in PAW were the reaction between NO (aq.) and OH (aq.), and NO (aq.) and $NO_2$ (aq.) with $H_2O$ (aq.). Similarly, the reaction that results in the formation of $NO_3^-$ ions (equations (22,31,36,37)) in PAW were given as the reaction between dissolved $O_3$ and $NO_2^-$ ions, and $NO_2$ and $H_2O$ (aq.), $NO_2^-$ and OH (aq.), and dissociation of peroxynitric acid (aq. ONOOH). The dissolution of gases $O_3$ in water results in the formation of dissolved $O_3$ (aq.). This was due to the particle solubility of $O_3$ in water at room temperature. The formation of $H_2O_2$ (equations (23,26,34)) in PAW occurs due to the reaction between OH molecules, $HO_2$ molecules, and $HO_2$ molecule and $O_2^-$ ion in water.

Along with the formation of RONS in PAW, the degradation of RONS also occurs in PAW to form more stable species in PAW. The dissolved $O_3$ and $NO_2^-$ ions react to form more stable $NO_3^-$ ions in PAW (equation (22)). Similarly, aqueous OH reacts with $NO_2^-$ ions reacts to form $NO_3^-$ ions in PAW in the acidic region (equation 36). Also, the $NO_2^-$ ions react with $H_2O_2$ to form peroxynitric acid, which is degraded to form $NO_3^-$ ions in PAW (equation (35,37)).

As discussed in Table 2, the formation of various reactive oxygen-nitrogen species in PAW. These species in PAW give PAW an immense potential to be used in various applications such as microbial (bacteria, fungi, virus, and pest, etc.) inactivation, food preservation, selective killing of cancer cells, and seeds germination and plant growth, etc.[1, 4-6, 14-17] The affinity of PAW for microbial inactivation and selective killing of cancer cells is due to the presence of various strong oxidizing species such as $H_2O_2$, dissolved $O_3$, hydroxyl radical (OH), peroxynitrile ($ONOO^-$), and superoxide ions ($O_2^-$), etc. as shown in Table 2[2, 5, 6, 11, 12, 14]. Along with strong oxidizing species, PAW is also a rich source of nitrogen species ($NO_3^-$, $NO_2^-$, etc.) which signifies the usefulness of PAW in the agriculture field[1, 4, 15-17].

Table 2 Mechanism of formation of reactive oxygen-nitrogen species in PAW[10-12, 21, 35-41]

| Reaction phase | Reaction | Rate constant or reaction rate | Equation number |
|---|---|---|---|
| Plasma phase | $N_2 + e^- \rightarrow 2N + e^-$ | $6.3 \times 10^{-6}\ T_e^{-1.6}\ e^{-9.8/T_e}$ cm$^3$ s$^{-1}$ | 8 |
| | $e^- + N_2^+ \rightarrow N + N^*$ | $2.0 \times 10^{-7}\ (T_e/0.03)^{-0.39}$ cm$^3$ s$^{-1}$ | 9 |
| | $H_2O + e^- \rightarrow OH + H + e^-$ | $2.6 \times 10^{-12}$ cm$^3$ s$^{-1}$ | 10 |
| | $H_2O + N_2(A) \rightarrow OH + H + N_2$ | $4.2 \times 10^{-11}$ cm$^3$ s$^{-1}$ | 11 |
| | $O_2 + e^- \rightarrow O + O + e^-$ | $3.2 \times 10^{-11}$ cm$^3$ s$^{-1}$ | 12 |
| | $N_2 + O \rightarrow NO + N$ | $2.7 \times 10^{-11}$ cm$^3$ s$^{-1}$ | 13 |
| | $N + O_2 \rightarrow NO + O$ | $8.5 \times 10^{-17}$ cm$^3$ s$^{-1}$ | 14 |
| | $N + OH \rightarrow H + NO$ | $4.7 \times 10^{-11}$ cm$^3$ s$^{-1}$ | 15 |
| | $O + NO + M \rightarrow NO_2 + M$ | $9.0 \times 10^{-32}$ cm$^6$ s$^{-1}$ | 16 |
| | $O + O_2 \rightarrow O_3$ | $1.7 \times 10^{-12}$ cm$^3$ s$^{-1}$ | 17 |
| | $OH + OH \rightarrow H_2O_2$ | $2.6 \times 10^{-11}$ cm$^3$ s$^{-1}$ | 18 |
| | $OH + O_3 \rightarrow HO_2 + O_2$ | $1.9 \times 10^{-12}$ cm$^3$ s$^{-1}$ | 19 |
| | $NO + O_3 \rightarrow NO_2 + O_2$ | $5.1 \times 10^{-12}$ cm$^3$ s$^{-1}$ | 20 |
| Plasma-liquid interphase | $N_2$, $e^-$, N, $N^*$, $H_2O$, OH, H, $O_2$, NO, $O_3$, $NO_2$, H, $HO_2$, $H_2O_2$, etc. | | |
| Liquid phase | $OH + NO \rightarrow \mathbf{NO_2^-} + H^+$ | $1.0 \times 10^{10}$ M$^{-1}$ s$^{-1}$ | 21 |
| | $\mathbf{O_3} + \mathbf{NO_2^-} \rightarrow O_2 + \mathbf{NO_3^-}$ | $2.5 \times 10^{5}$ M$^{-1}$ s$^{-1}$ | 22 |
| | $OH + OH \rightarrow \mathbf{H_2O_2}$ | $5.0 \times 10^{9}$ M$^{-1}$ s$^{-1}$ | 23 |
| | $OH + \mathbf{O_3} \rightarrow O_2 + HO_2$ | $1.0 \times 10^{8}$ M$^{-1}$ s$^{-1}$ | 24 |
| | $OH + HO_2 \rightarrow O_2 + H_2O$ | $7.5 \times 10^{9}$ M$^{-1}$ s$^{-1}$ | 25 |
| | $HO_2 + HO_2 \rightarrow O_2 + \mathbf{H_2O_2}$ | $1.0 \times 10^{6}$ M$^{-1}$ s$^{-1}$ | 26 |
| | $HO_2 + NO \rightarrow ONOO^- + H^+$ | $3.2 \times 10^{9}$ M$^{-1}$ s$^{-1}$ | 27 |
| | $OH + NO_2 \rightarrow ONOO^- + H^+$ | $1.2 \times 10^{10}$ M$^{-1}$ s$^{-1}$ | 28 |
| | $NO + NO + O_2 \rightarrow 2NO_2$ | $2.3 \times 10^{6}$ M$^{-2}$ s$^{-1}$ | 29 |
| | $NO_2 + NO + H_2O \rightarrow \mathbf{2NO_2^-} + 2H^+$ | $2.0 \times 10^{8}$ M$^{-1}$ s$^{-1}$ | 30 |
| | $2NO_2 + H_2O \rightarrow \mathbf{NO_3^-} + \mathbf{NO_2^-} + 2H^+$ | $0.5 \times 10^{8}$ M$^{-1}$ s$^{-1}$ | 31 |
| | $\mathbf{H_2O_2} + OH \rightarrow H_2O + O_2^- + H^+$ | $2.7 \times 10^{7}$ M$^{-1}$ s$^{-1}$ | 32 |
| | $O_2^- + NO \rightarrow ONOO^-$ | $5.0 \times 10^{9}$ M$^{-1}$ s$^{-1}$ | 33 |
| | $H_2O + HO_2 + O_2^- \rightarrow O_2 + \boldsymbol{H_2O_2} + HO^-$ | 9.7 × 10$^7$ M$^{-1}$ s$^{-1}$ | 34 |
| | $\boldsymbol{NO_2^-} + \boldsymbol{H_2O_2} + H^+ \rightarrow ONOOH + H_2O$ | 1.1 × 10$^3$ M$^{-1}$ s$^{-1}$ | 35 |
| | $\boldsymbol{NO_2^-} + OH + H^+ \rightarrow \boldsymbol{NO_3^-} + 2H^+$ | 5.3 × 10$^9$ M$^{-1}$ s$^{-1}$ | 36 |
| | $ONOOH \rightarrow \boldsymbol{NO_3^-} + H^+$ | 0.9 s$^{-1}$ | 37 |

### 3.3 The effect of knurling on the physicochemical properties of PAW and RONS concentration

The variation in the physicochemical properties of PAW and reactive oxygen-nitrogen species (RONS) concentration while varying knurling pitch size is shown in figure 5. Figure 5 showed that introducing knurling to the ground electrode significantly ($p < 0.05$) improves the physicochemical properties of PAW and RONS concentration in it. This signifies the increase in species/radicals produced in the plasma phase when the ground electrode has knurling compared to the non-knurled electrode. This behavior was also observed in the electric characterization of plasma. In which, higher discharge current and power dissipation were observed in plasma with ground electrode knurling compared to the non-knurled electrode.

The plasma-water interaction decreased the pH of PAW due to the formation of various acidic species in water. The most leading acidic species are $NO_2^-$ ions and $NO_3^-$ ions in the form of nitrous and nitric acid. Increasing plasma-water treatment time results in the formation of more acidic species in PAW as a result, we observed a continuous decrease in the pH of PAW as shown in figure 5 (a). Moreover, the pH of PAW prepared using plasma device with knurling and without knurling have significant ($p < 0.05$) difference among them. With knurling (2 mm knurling pitch), the pH of PAW decreased by 53.0% after 5 min of plasma-water treatment compared to control. However, without knurling pH of PAW decreased by 25.8% after 5 min of plasma-water treatment compared to control only. In addition, increasing knurling pitch from 0.5 mm to 2 mm showed a decrease in pH of PAW, however, this decrease in pH of PAW was not statistically significant ($p < 0.05$). Hence, introducing knurling to the ground electrode has a substantial impact on the pH of PAW. Since, as discussed in figure 3 (a), introducing knurling increased the plasma discharge and resulted in the formation of more acidic plasma species that dissolved in water and decreased the pH of PAW.

The effect of knurling on oxidation-reduction potential (ORP) of PAW is shown in figure 5 (b). The ORP gives the oxidizing tendency of PAW which could be used as an indicator of the antimicrobial activity of PAW. Since, it gives the net combination of all oxidizing species (dissolved $O_3$, $H_2O_2$, $^{\cdot}OH$, free electrons, etc.) present in PAW. Increasing plasma-water treatment time increased the ORP of PAW which showed the increase in dissolution of oxidizing species in PAW. The knurling of the ground electrode also helps in increasing the oxidizing tendency of PAW as shown in figure 5 (b). The ORP of PAW prepared using a 1 mm knurling ground electrode (ORP – 600 mV) was 18.8% higher compared to without a knurling ground electrode (ORP – 505 mV). Moreover, the ORP of PAW prepared using a 1 mm knurling electrode was substantially ($p < 0.05$) higher compared to 0.5 mm and 2 mm knurling electrodes. This was due to higher power dissipation in the 1 mm knurling electrode (figure 3 (a)) resulting in the formation of more oxidizing species in PAW.

The net combination of all inorganic ions ($H^+$, $NO_3^-$ ions, $NO_2^-$ ions, $OONO^-$, etc.) present in PAW were measured using total dissolved solids (TDS) and electrical conductivity (EC). Figure 5 (c, d) showed the variation in TDS and EC of PAW while varying knurling pitch and plasma-water treatment time. Increasing plasma-water exposure time significantly ($p < 0.05$) increases the TDS and EC of PAW showing the continuous formation of inorganic ions in PAW. In addition, introducing knurling to the ground electrode substantially ($p < 0.05$) increased the TDS and EC of PAW. The TDS and EC of PAW with ground electrode knurling (0.5 mm knurling pitch) were 616.7% and 567.7% higher than without ground electrode knurling. Initially, the TDS and EC of PAW were prepared using a 0.5 mm knurling electrode plasma device that showed a higher value compared to other knurling pitches. As the plasma-water treatment time increases, this difference in TDS and EC of PAW keeps on decreasing and becomes statistically insignificant ($p > 0.05$) at 5 min.

The variation in $NO_3^-$ ions and $NO_2^-$ ions concentration (reactive nitrogen species) in PAW with varying ground electrode knurling pitch and plasma-water treatment time is shown in figure 5 (e, f). Increasing plasma-water treatment continuously increased the $NO_3^-$ ions and $NO_2^-$ ions concentration in PAW for all knurling pitch ranges. Moreover, the knurling of the ground electrode significantly increased the $NO_3^-$ ions and $NO_2^-$ ions concentration in PAW compared to a non-knurled ground electrode. The maximum concentration of $NO_3^-$ ions and $NO_2^-$ ions in PAW (plasma-water treatment time of 5 min) with and without knurling ground electrode were given as 24.98 mg $l^{-1}$ and 5.21 mg $l^{-1}$ $NO_3^-$ ions and 2.80 mg $l^{-1}$ and 1.11 mg $l^{-1}$ $NO_2^-$ ions, respectively. Moreover, the larger knurling pitch size (2 mm) did not support the higher formation of $NO_3^-$ ion concentration compared to the smaller knurling pitch size (0.5 mm and 1 mm). Hence, considering the appropriate knurling pitch size is also important to get higher production of $NO_3^-$ ions concentration in PAW. For $NO_2^-$ ions, 1 mm knurling pitch gives the highest concentration of $NO_2^-$ ions in PAW compared to 0.5 mm and 2 mm knurling pitch.

The variation in reactive oxygen species (ROS ($H_2O_2$ and dissolved $O_3$)) with plasma-water treatment time and ground electrode knurling pitch size is shown in figure 5 (g, h). As the plasma-water treatment time increases, the continuous increase in the $H_2O_2$ and dissolved $O_3$ concentration were observed in PAW when the ground electrode without knurling was used. However, the $H_2O_2$ and dissolved $O_3$ concentration in PAW when prepared using a ground electrode with knurling showed an initial increase and then decreased with time with an exception of $H_2O_2$ at 5 min of plasma-water treatment time. This was due to the reactivity environment favoring the reaction within ROS and ROS reaction with $NO_2^-$ ions to give more stable $NO_3^-$ ions (equations (22,24,32,35-37)). Thus, the concentration of $NO_3^-$ ions in PAW is substantially higher compared to other RONS in PAW. Hence, the ROS concentration decreased at a higher plasma-water treatment time.

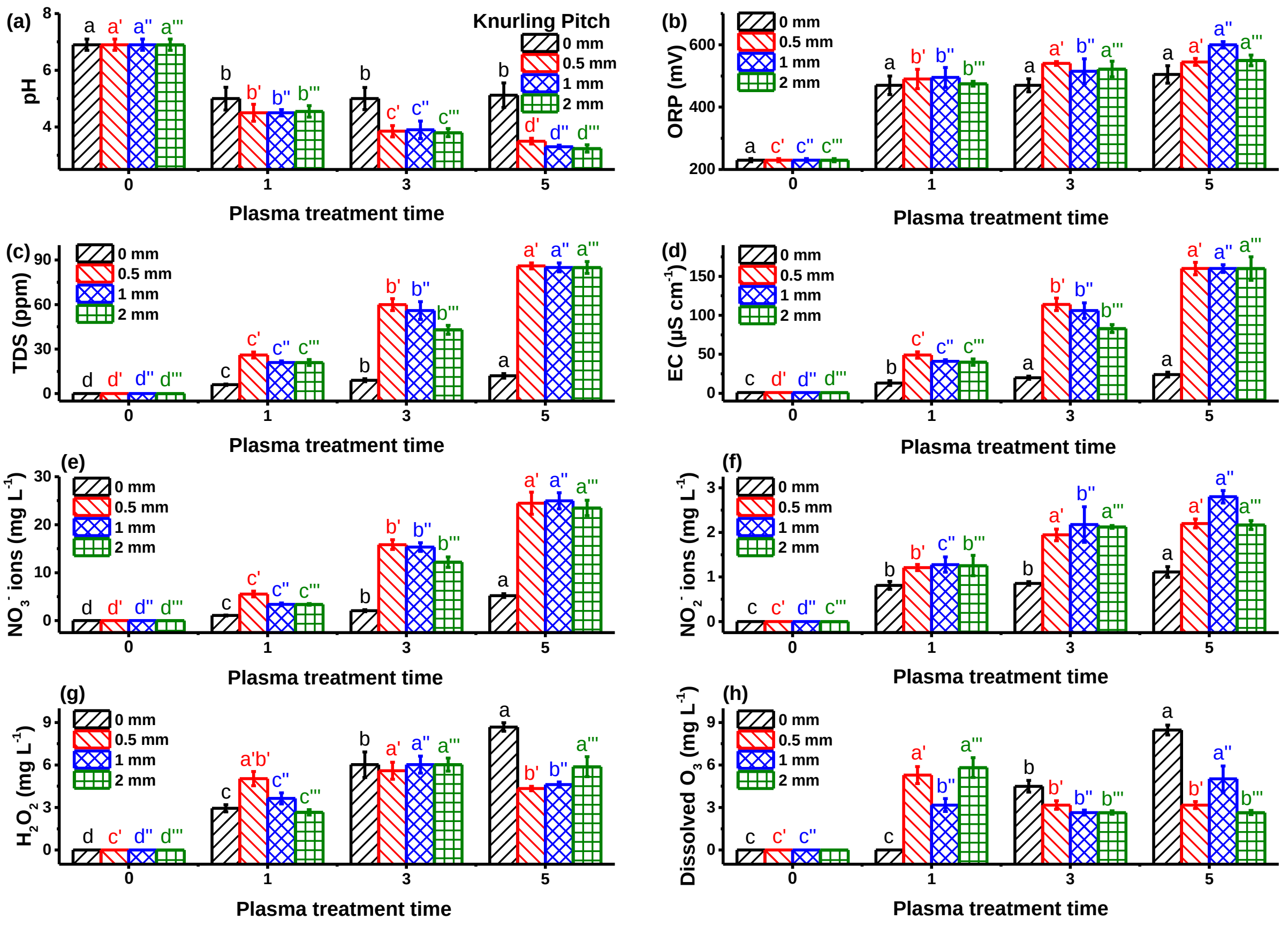

Figure 5. The variation in physicochemical properties of PAW and RONS concentration with varying ground electrode knurling pitch (0 mm, 0.5 mm, 1 mm, and 2 mm). Different lowercase letters showed statistically significant difference ($p<0.05$, $n \geq 3$) among the properties of PAW mean ± standard deviation ($\mu \pm \sigma$)

3.4 The effect of ground electrode material on the physicochemical properties of PAW and RONS concentration

The role of ground electrode material of plasma device on the physicochemical properties and RONS concentration of PAW is shown in figure 6. The observed values of physicochemical properties of PAW when copper (Cu) was used as a ground electrode material was higher compared to stainless steel (SS) and brass (Br). This can be implied from the plasma discharge power in which Cu had the higher plasma discharge power compared to Br and SS (figure 3 (b)). Higher discharge power signifies more generation of plasma radicals/species which come in contact with water and improved the physicochemical properties of PAW and RONS concentration in it. The lowest pH of PAW (figure 6 (a)) PAW (5 min of plasma-water treatment time) when prepared using Cu, Br, and SS as ground electrode material were given as 3.1, 3.56, and 3.5, respectively. This was also reflected as higher TDS, EC, $NO_2^-$ + $NO_3^-$ ions concentration (figure 6 (c-f)) in PAW prepared using Cu as ground electrode compared to Br and SS. The TDS and EC of PAW when prepared using Cu as ground electrode (5 min of plasma-water treatment time) were 12.8% and 12.5% higher than SS, and 29.3% and 24.1% higher than Br, respectively. Similarly, for a plasma-water treatment time of 5 min, the observed $NO_3^-$ ions concentration in PAW when prepared using Cu as ground electrode was 14.3% and 27.1% higher than SS and Br.

The oxidizing potential (ORP) of PAW prepared using Cu and SS as the ground electrode was slightly higher than Br. The ORP of PAW prepared using Cu, SS, and Br as ground electrodes was given as 548 mV, 545 mV, and 535 mV, respectively. The slightly higher ORP and lower pH of PAW prepared using Cu and SS as ground electrodes compared to Br increases PAW reactivity. The reactive environment of PAW favors reactions within ROS and generated ROS with $NO_2^-$ ions. As a result, the concentration of $NO_2^-$ ions and ROS decreased in the high reactive environment of PAW. The results of $NO_2^-$ ions concentration in PAW shown in figure 6 (f) confirm the same. Initial for a plasma-water treatment time of 1 min, the observed $NO_2^-$ ions concentration prepared using Br as ground electrode lower than Cu and SS. However, as the plasma-water treatment time increased (3 min and 5 min), the observed $NO_2^-$ ions concentration in PAW prepared using Br as the ground electrode was higher than Cu and SS. Since, increasing plasma-water treatment time, increased the reactivity of PAW as a result $NO_2^-$ ions present in PAW react with dissolved $O_3$ and $H_2O_2$. Therefore, decreased the concentration of $NO_2^-$ ions in PAW prepared using Cu and SS as ground electrodes material compared to Br. Similar results were observed in the ROS concentration in PAW as shown in figure 6 (g, h) with an exception of $H_2O_2$ in PAW (5 min treatment time) prepared using Cu as a ground electrode. In which a high reactive PAW, the $H_2O_2$ and dissolved $O_3$ concentration present in PAW either decrease or remain constant. As discussed, the PAW prepared using Br as ground electrode material had comparatively low reactivity, hence, the increase in the concentration of $H_2O_2$ and dissolved $O_3$ were observed with time. Moreover, the concentration of $H_2O_2$ and dissolved $O_3$ in PAW decreases or remains constant as the plasma-water treatment time increases (or, the reactivity of PAW increases) for SS and Cu as ground electrode material. One exception was also observed in $H_2O_2$ concentration in PAW (figure 6 (g)), in which a higher concentration of $H_2O_2$ was observed in high reactivity PAW when prepared using Cu as ground electrode material. The possible reason for the same is an excess concentration of $H_2O_2$ in PAW which is left even after reaction with dissolved $O_3$ and $NO_2^-$ ions present in PAW.

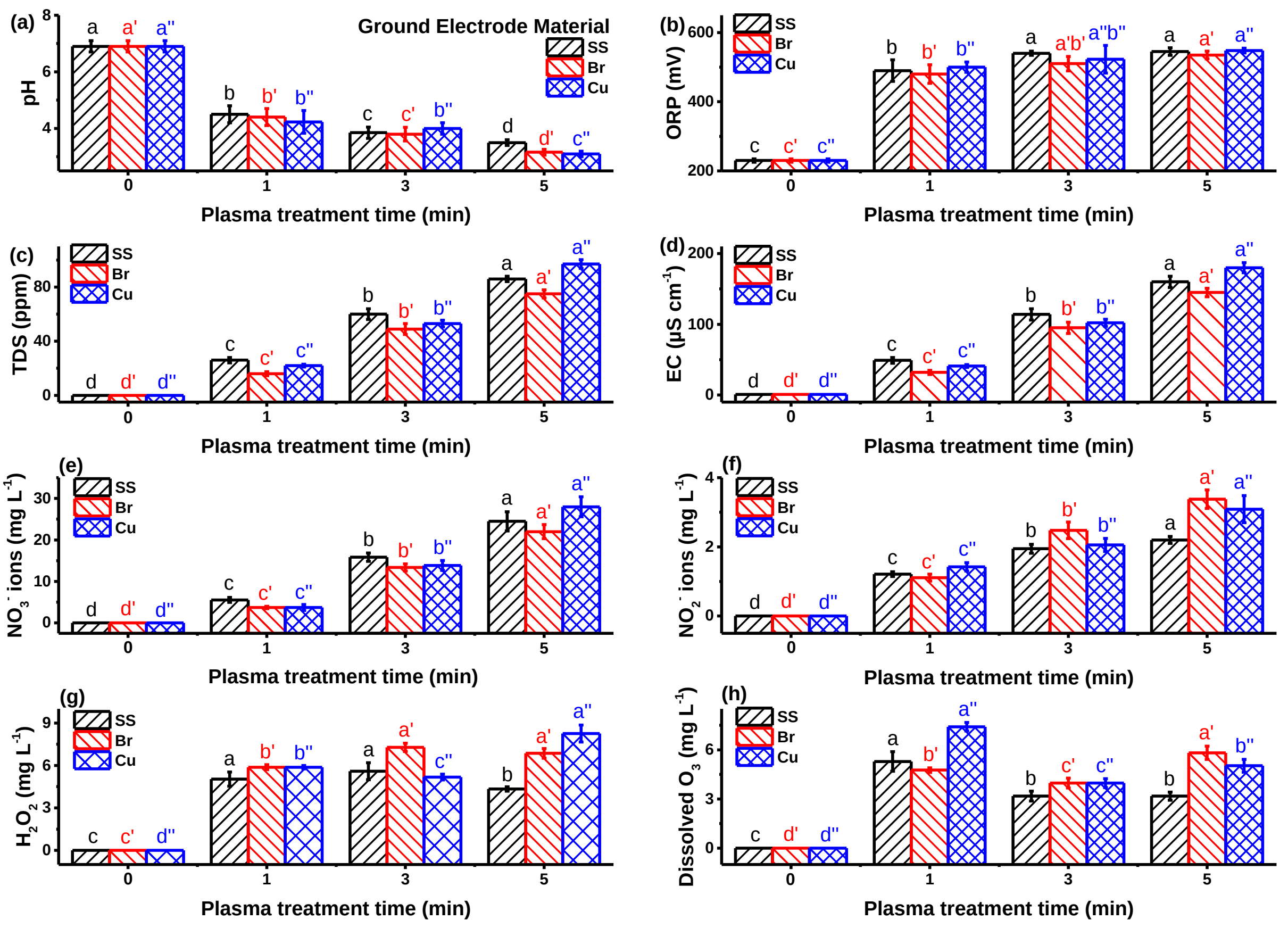

Figure 6. The variation in physicochemical properties of PAW and RONS concentration with varying ground electrode material (SS, Br, and Cu). Different lowercase letters showed statistically significant difference ($p<0.05$, $n \geq 3$) among the properties of PAW mean ± standard deviation ($\mu \pm \sigma$)

## 3.5 The effect of power electrode material on the physicochemical properties of PAW and RONS concentration

The variation in the power electrode material on the physicochemical properties of PAW and RONS concentration is shown in figure 7. Similar to ground electrode material, the use of Cu for power electrode material significantly enhances the physicochemical properties of PAW and RONS concentration in it. The results also supported the plasma discharge power in which Cu had substantially higher power compared to Br and SS (figure 3 (e)). The lowest pH of PAW (5 min of plasma treatment time) when prepared using Cu, Br, and SS as power electrode material (figure 7 (a)) were given as 2.9, 3, and 3.3, respectively. Hence, the PAW produced using Cu as power electrode material generates PAW with higher acidity compared to other materials.

At a higher plasma-water treatment time, the oxidizing potential (ORP) of PAW prepared using Br and Cu as power electrode material was significantly ($p < 0.05$) higher compared to SS (figure 7 (b)). In addition, the ORP of PAW prepared using Cu (ORP – 590 mV) as a power electrode was slightly higher compared to Br (ORP – 580 mV). Hence, the generation of oxidizing species in PAW during plasma-water exposure was substantially higher when Cu or Br was used as power electrode material.

The significance of Cu material as a power electrode over SS and Br is easily understood by comparing the results of TDS, EC, $NO_3^-$ ions, and $NO_2^-$ ions concentration in PAW. For a plasma-water treatment time of 5 min, a substantial growth ($p < 0.05$) in the values of TDS, EC, $NO_3^-$ ions, and $NO_2^-$ ions in PAW when prepared using Cu as power electrode material compared to SS and Br as shown in figure 7 (c-f). The increase in TDS and EC of PAW using Cu as power electrode were 100.0% and 107.9% higher compared to Br, and 92.3% and 96.4% higher compared to SS. Similarly, the increase in $NO_3^-$ and $NO_2^-$ ions concentration in PAW prepared using Cu as power electrode material were 109.0% and 18.8% higher compared to Br, and 92.2% and 40.7% higher compared to SS. Hence, the use of Cu as an outer electrode material helps in the generation of more inorganic ions in PAW which was communicated above as enhancement in TDS, EC, $NO_3^-$ ions, and $NO_2^-$ ions concentration in PAW.

The variation in $H_2O_2$ and dissolved $O_3$ when using different power electrode materials is shown in figure 7 (g, h). The $H_2O_2$ and dissolved $O_3$ concentration present in PAW when using different materials for power electrode showed a rise and fall (or rise and remain constant) in its concentration with increasing plasma-water treatment time. Since, initially (t = 0 min), there was no $H_2O_2$ and dissolved $O_3$ present in PAW. Hence, as soon as plasma-water interaction starts occurring, the formation of $H_2O_2$ and dissolved $O_3$, as a result, their concentration starts increasing. The fall in concentration of $H_2O_2$ and dissolved $O_3$ showed (SS power electrode) reduction of these species due to reaction occurring among themselves and with $NO_2^-$ ions to form more stable $NO_3^-$ ions. Moreover, the constant concentration of $H_2O_2$ and dissolved $O_3$ (Br and Cu power electrode) with increasing time showed the established equilibrium in which excess concentration above the equilibrium point converted more stable products like $NO_3^-$ ions.

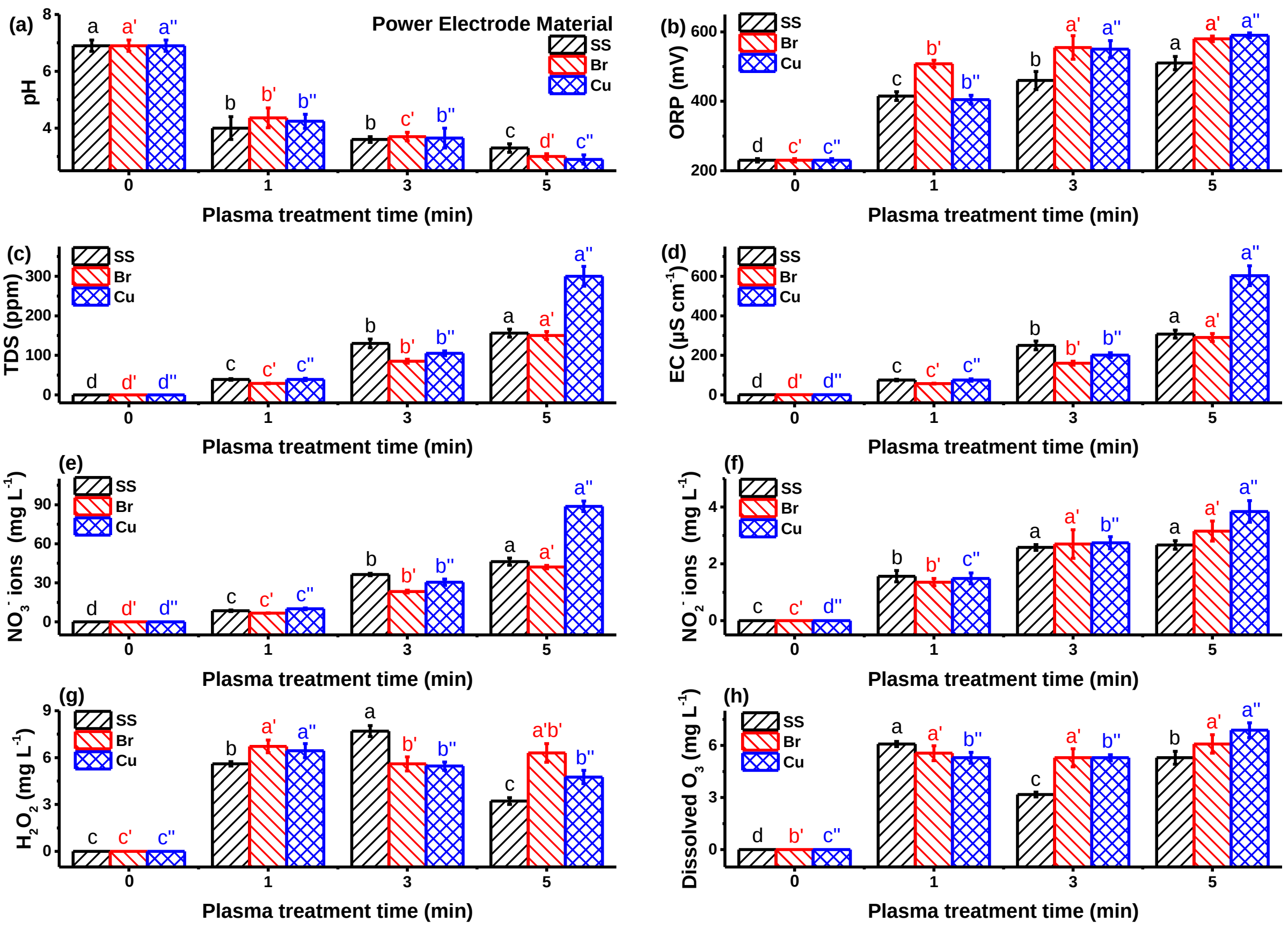

Figure 7. The variation in physicochemical properties of PAW and RONS concentration with varying power electrode material (SS, Br, and Cu). Different lowercase letters showed

statistically significant difference ($p<0.05$, $n \geq 3$) among the properties of PAW mean ± standard deviation ($\mu \pm \sigma$)

3.6 The effect of power electrode type on the physicochemical properties of PAW and RONS concentration

In this section, we discussed the role of different types of electrodes on the physicochemical properties of PAW and RONS concentration. For this study, three different types of electrodes were used such as mesh, wire, and sheet. The observed result is shown in figure 8. The results of physicochemical properties of PAW (figure 8 (a-d)) showed the use of wire as a power electrode type significantly enhanced the physicochemical properties of PAW compared to other electrode types. This was due to the use of wire as a power electrode having higher plasma discharge power compared to mesh and sheet (figure 3 (c)). The pH of PAW, when prepared using wire as a power electrode (5 min of plasma treatment time) showed 11.4% lower compared to mesh, and 7.7% lower compared to the sheet. Similarly, the oxidizing potential (ORP) of PAW when prepared using wire as a power electrode was 7.2% higher than mesh and 5.6% higher compared to the sheet. Moreover, at the same operating parameters, the TDS and EC of PAW prepared using wire as power electrode showed 19.4% and 17.6% higher compared to mesh, and 32.3% and 30.0% higher compared to the sheet.

Similar to physicochemical properties, the $NO_3^-$ ions concentration present in PAW prepare (5 min of plasma treatment time) using wire as power electrode type showed significantly higher value compared to mesh and sheet (figure 8 (e)). The observed $NO_3^-$ ions concentration in PAW prepared using wire, mesh, and sheet as power electrodes were given as 27.1 mg $l^{-1}$, 24.5 mg $l^{-1}$, and 19.8 mg $l^{-1}$, respectively.

In contrast to physicochemical properties and $NO_3^-$ ions concentration, the $NO_2^-$ ions, $H_2O_2$, and dissolved $O_3$ present in PAW showed higher value for the sheet as power compared to wire and mesh (figure 8 (f-h)). This signifies the PAW produced using the sheet as a power electrode did not favors the reaction within dissolved ROS ($H_2O_2$ and dissolved $O_3$) and with $NO_2^-$ ions. Hence, due to limiting reactions (equations (22, 35-37)) between these species, the observed concentration of reactants ($H_2O_2$, dissolved $O_3$, and $NO_2^-$ ions) were high and the product concentration ($NO_3^-$ ions) was low.

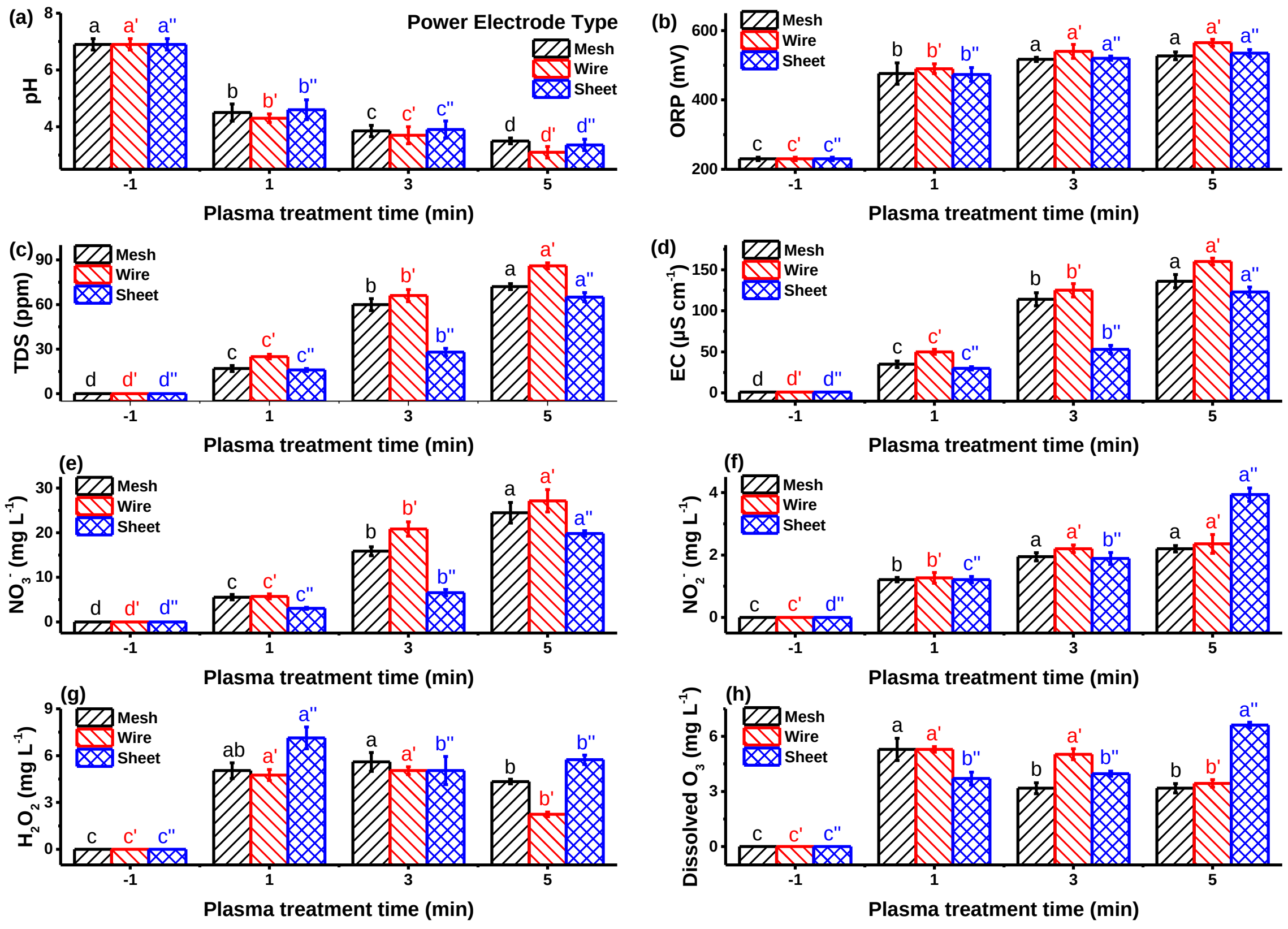

Figure 8. The variation in physicochemical properties of PAW and RONS concentration with varying power electrode types (mesh, wire, and sheet). Different lowercase letters showed statistically significant difference ($p<0.05$, $n \geq 3$) among the properties of PAW mean ± standard deviation ($\mu \pm \sigma$)

3.7 The effect of dielectric material on the physicochemical properties of PAW and RONS concentration

Figure 9 showed the variation in physicochemical properties of PAW and RONS concentration when using glass and quartz dielectric materials in plasma device. The results showed statistically significant ($p < 0.05$) enhancement in the physicochemical properties of PAW and RONS concentration when quartz was used as a dielectric material compared to glass. As discussed in the electrical characterization, higher discharge current peaks and plasma discharge power in quartz as dielectric compared to glass (figures 2 (c, d) and 3 (d)) [25]. This signifies higher plasma species/radicals density in quartz plasma device compared to glass. As a result, a high concentration of reactive species dissolved in PAW was prepared using a quartz plasma device as shown by enhanced physicochemical properties and RONS concentration compared to a glass plasma device. The ORP, TDS, and EC of PAW when prepared using quartz as dielectric showed a higher value and lower value of pH of PAW compared to glass (figure 9 (a-d)). This showed the improvement in the physicochemical properties of PAW when prepared using quartz. The pH of PAW (plasma treatment time of 5 min) when prepared using glass and quartz as dielectric were given as 3.5 and 3. At similar conditions, the ORP, TDS, and EC of PAW when prepared using glass and quartz were given as 545 mV and 580 mV (ORP), 86 ppm and 150 ppm (TDS), and 160 $\mu S\ cm^{-1}$ and 290 $\mu S\ cm^{-1}$ (EC), respectively.

The variation in dissolved RONS in PAW when prepared using glass and quartz as the dielectric material of the plasma device is shown in figure 9 (e-h). The observed $NO_3^-$ and $NO_2^-$ ions concentration in PAW prepared using quartz as dielectric significantly ($p < 0.05$) higher compared to glass (figure 9 (e-f)). At a plasma-water treatment time of 5 min, the $NO_3^-$ and $NO_2^-$ ions concentration in PAW were 71.8% and 45.4% higher compared to glass. Similar to $NO_3^-$ and $NO_2^-$ ions, the observed $H_2O_2$ and dissolved $O_3$ concentrations in PAW were higher for quartz compared to glass. For glass, with increasing plasma-water treatment time, the H2O2 concentration and dissolved O3 in PAW showed a rise and fall. This was as the reactivity of PAW increases these species react with each other and other NO2- ions present in PAW. As a result, their concentration decreased at higher plasma-water treatment time. For quartz, the $H_2O_2$ and dissolved $O_3$ present in PAW follow a trend of rise-fall-rise with time (figure 9 (g, h)). This was due to initially no $H_2O_2$ and dissolved $O_3$ present in PAW. Hence, these species' concentration increases with increasing time. As the concentration of these reached sufficient reactivity, they react with each other, and $NO_2^-$ ions concentration was present in PAW. As a result, a decrease in their concentration was observed. Further increase in the plasma-water treatment time showed a generation of a higher concentration of ROS in PAW. Hence, even after reacting with each other and $NO_2^-$ ions present in PAW a slight increase in these species ($H_2O_2$ and dissolved $O_3$) concentration was observed in PAW.

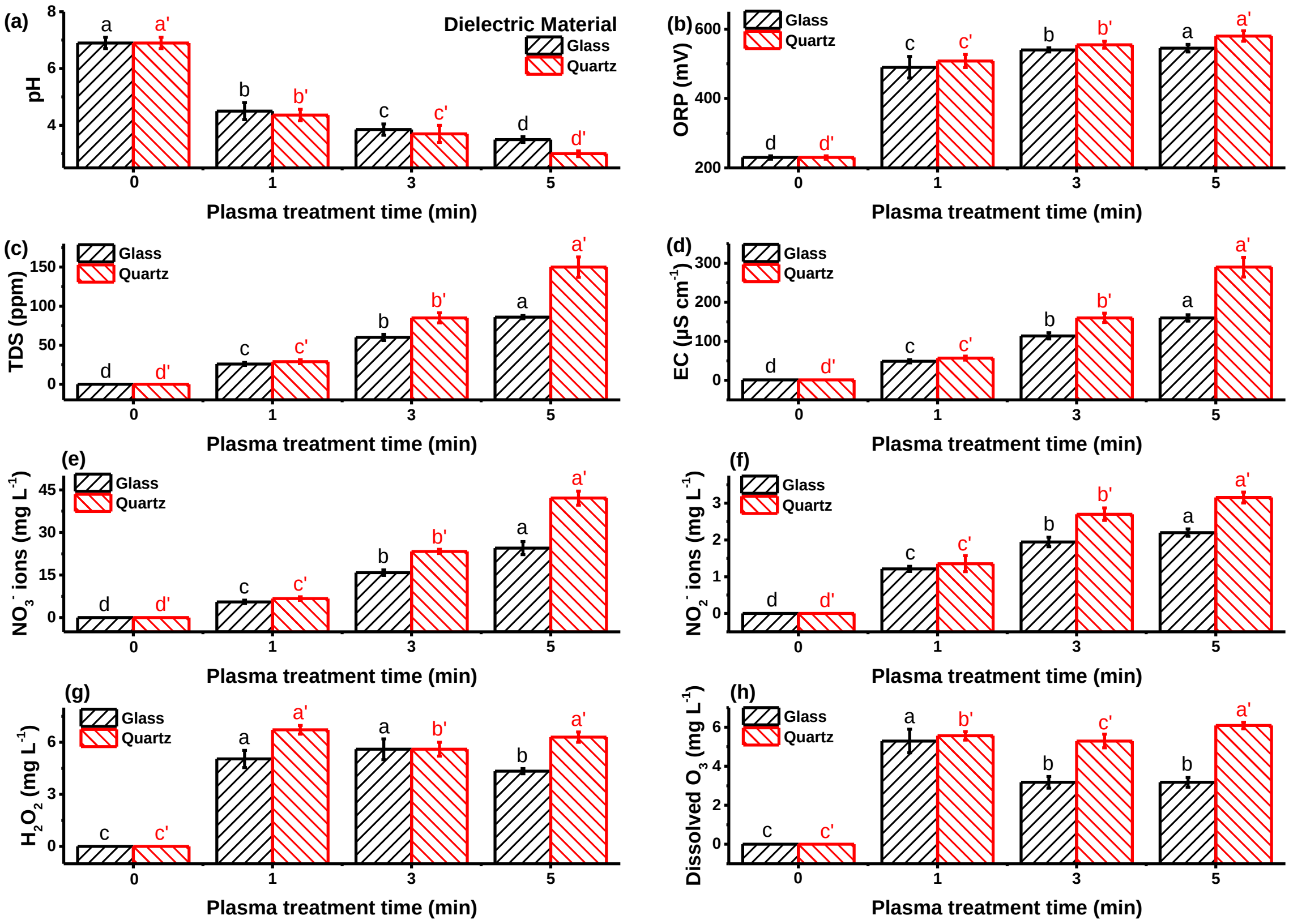

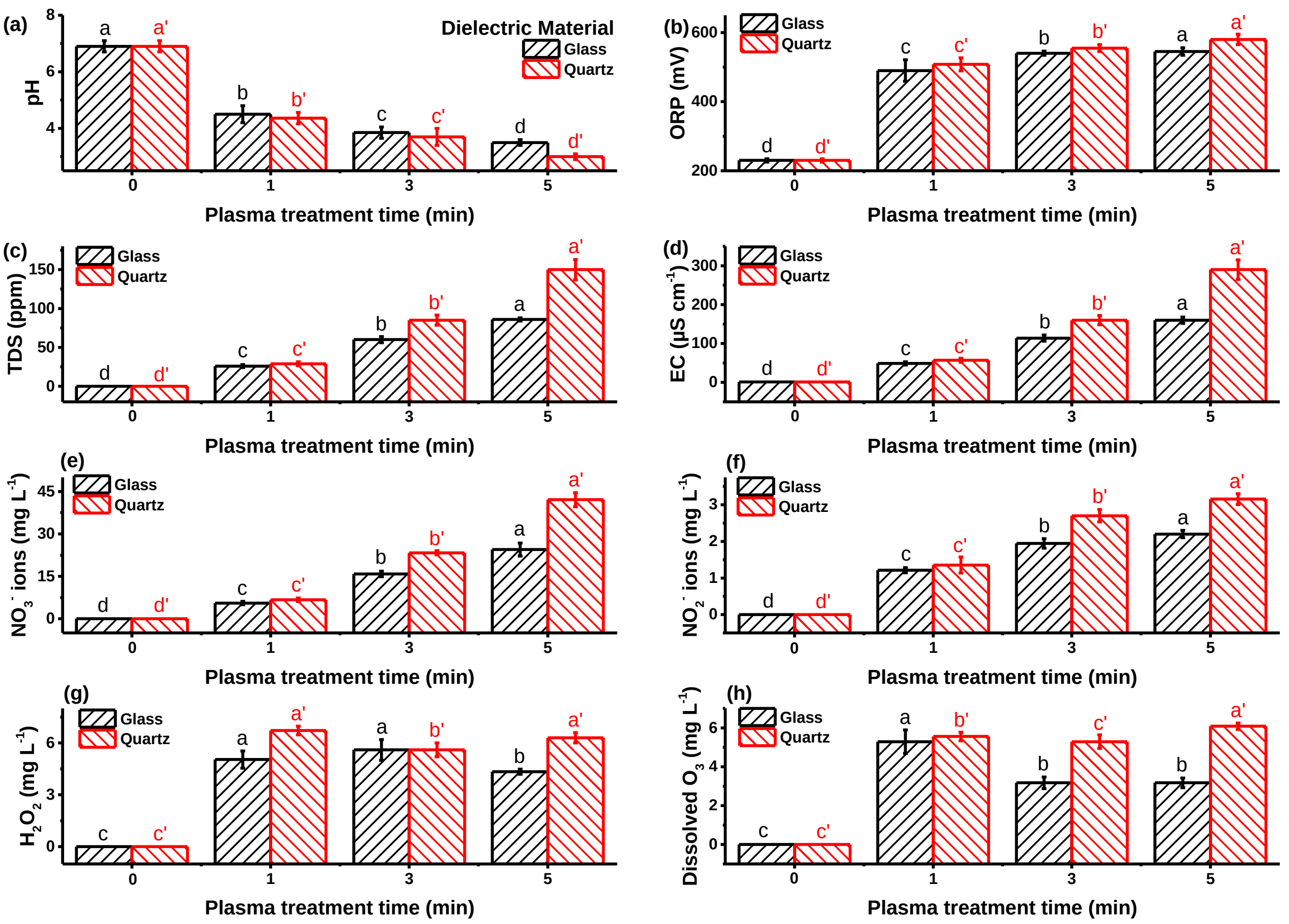

Figure 9. The variation in physicochemical properties of PAW and RONS concentration with varying dielectric material (glass and quartz). Different lowercase letters showed statistically significant difference ($p<0.05$, $n \geq 3$) among the properties of PAW mean ± standard deviation ($\mu \pm \sigma$)

## 3.8 Residual metal analysis in PAW

As PAW has various applications in the field of agriculture, food preservation, and cancer cells inactivation, etc. The presence of heavy metal in PAW coming from electron erosion used in plasma device may interfere in the applications of PAW. Hence, the residual metal analysis of PAW become important. Table 1 shows the residual metal in PAW when prepared using different ground electrode material. The different material of construction of ground electrode are stainless steel (SS, allow of iron (Fe) and carbon), brass (Br, copper and zinc (Zn)), and copper (Cu). Hence, the Table 1 showed the concentration of Cu, Zn, and Fe in PAW and control. The observed concentration of Cu and Zn present in PAW and control were insignificant ($p > 0.05$) for different ground electrode material of plasma device. Moreover, the concentration of Cu and Zn were less than 6 µg $l^{-1}$ and 2 µg $l^{-1}$. Also, the Fe concentration in PAW and control was beyond the detection limit (B.D.L). In conclusion, the observed heavy metal concentration in PAW similar to ultrapure milli-Q water (control). Hence, no erosion of inner (ground) electrode material occurs during plasma-water treatment which settle in PAW in form of residue metal.

Table 1. Residual metal analysis in PAW when prepared using different ground electrode material. Different lowercase letters showed statistically significant difference ($p<0.05$, $n \geq 3$) among the group mean ± standard deviation ($\mu \pm \sigma$) (*B.D.L – Below Detection Limit)

| Material of ground electrode | Plasma-water treatment time | Residue metal present in water | | |
|---|---|---|---|---|
| | | Copper (µg $l^{-1}$) | Zinc (µg $l^{-1}$) | Iron (µg $l^{-1}$) |
| Control | 0 min | $3.9 \pm 0.2^{a}$ | $1.6 \pm 0.14^{a}$ | B.D.L* |
| Stainless Steel | 1 min | $4.6 \pm 1.1^{a}$ | $1.3 \pm 0.3^{a}$ | B.D.L* |
| | 3 min | $3.9 \pm 0.8^{a}$ | $1.7 \pm 0.5^{a}$ | B.D.L* |
| | 5 min | $4.7 \pm 0.7^{a}$ | $1.8 \pm 0.4^{a}$ | B.D.L* |
| Brass | 1 min | $4.5 \pm 0.5^{a}$ | $1.7 \pm 0.2^{a}$ | B.D.L* |
| | 3 min | $4.9 \pm 1.2^{a}$ | $1.2 \pm 0.3^{a}$ | B.D.L* |
| | 5 min | $5.0 \pm 1.0^{a}$ | $1.7 \pm 0.2^{a}$ | B.D.L* |
| Copper | 1 min | $5.1 \pm 0.9^{a}$ | $1.6 \pm 0.1^{a}$ | B.D.L* |
| | 3 min | $4.8 \pm 0.5^{a}$ | $1.6 \pm 0.3^{a}$ | B.D.L* |
| | 5 min | $5.8 \pm 0.7^{a}$ | $1.7 \pm 0.5^{a}$ | B.D.L* |

## 4. Conclusion

The present work shows the effect of various components of dielectric barrier discharge plasma device (DBD-PD) on plasma and physicochemical properties of PAW and RONS concentration present in activated water. The discharge current characteristics showed substantial improvement in filamentary discharge when introduced knurling to ground electrode, using quartz as dielectric layer compared to glass, and employed wired power electrode compared to mesh and sheet. These results are further supported by the results of plasma discharge power.

Similarly, the plasma activated water produced using a diamond knurled electrode, quartz as dielectric, and wire as power electrode showed substantial improvement in the physicochemical properties of PAW and RONS concentration. We have also observed that the use of copper material for manufacturing ground and power electrodes significantly improves the PAW properties compared to brass and stainless steel materials.

## Acknowledgments

This work was supported by the Department of Atomic Energy (Government of India) graduate fellowship scheme (DGFS). The authors sincerely thank Mr Chirayu Patil, O. R. Kaila, and Mr. Nimish for providing constant support and useful suggestions during this work.

## Data availability statement

The data that support the findings of this study are available upon reasonable request from the authors.

## Conflict of interests

The authors declare that there are no conflicts of interests.

**Authors' contributions**

Both authors contributed to the study conception and design. Material preparation, data collection, and analysis were performed by Vikas Rathore. The first draft of the manuscript was written by Vikas Rathore, and both authors commented on previous versions of the manuscript. Both authors read and approved the final manuscript.

**ORCID iDs**

Vikas Rathore https://orcid.org/0000-0001-6480-5009

**References**

[1] Thirumdas R, Kothakota A, Annapure U, Siliveru K, Blundell R, Gatt R, Valdramidis V P J T i f s and technology 2018 Plasma activated water (PAW): Chemistry, physico-chemical properties, applications in food and agriculture **77** 21-31

[2] Guo L, Xu R, Gou L, Liu Z, Zhao Y, Liu D, Zhang L, Chen H, Kong M G J A and microbiology e 2018 Mechanism of virus inactivation by cold atmospheric-pressure plasma and plasma-activated water **84** e00726-18

[3] Kaushik N K, Ghimire B, Li Y, Adhikari M, Veerana M, Kaushik N, Jha N, Adhikari B, Lee S-J and Masur K J B c 2019 Biological and medical applications of plasma-activated media, water and solutions **400** 39-62

[4] Rathore V, Tiwari B S, Nema S K J P C and Processing P 2022 Treatment of Pea Seeds with Plasma Activated Water to Enhance Germination, Plant Growth, and Plant Composition **42** 109-29

[5] Rathore V, Patel D, Butani S, Nema S K J P C and Processing P 2021 Investigation of physicochemical properties of plasma activated water and its bactericidal efficacy **41** 871-902

[6] Rathore V, Patel D, Shah N, Butani S, Pansuriya H, Nema S K J P C and Processing P 2021 Inactivation of Candida albicans and lemon (Citrus limon) spoilage fungi using plasma activated water **41** 1397-414

[7] Rathore V and Nema S K J J o A P 2021 Optimization of process parameters to generate plasma activated water and study of physicochemical properties of plasma activated solutions at optimum condition **129** 084901

[8] Rathore V and Nema S K J P S 2022 The role of different plasma forming gases on chemical species formed in plasma activated water (PAW) and their effect on its properties

[9] Rathore V and Nema S K J P o P 2022 A comparative study of dielectric barrier discharge plasma device and plasma jet to generate plasma activated water and post-discharge trapping of reactive species **29** 033510

[10] Rathore V, Patil C, Sanghariyat A and Nema S K J T E P J D 2022 Design and development of dielectric barrier discharge setup to form plasma-activated water and optimization of process parameters **76** 1-14

[11] Lukes P, Dolezalova E, Sisrova I, Clupek M J P S S and Technology 2014 Aqueous-phase chemistry and bactericidal effects from an air discharge plasma in contact with water: evidence for the formation of peroxynitrite through a pseudo-second-order post-discharge reaction of H2O2 and HNO2 **23** 015019

[12] Myers B, Ranieri P, Smirnova T, Hewitt P, Peterson D, Quesada M H, Lenker E and Stapelmann K J J o P D A P 2021 Measuring plasma-generated· OH and O atoms in liquid using EPR spectroscopy and the non-selectivity of the HTA assay **54** 145202

[13] Ten Bosch L, Köhler R, Ortmann R, Wieneke S, Viöl W J I j o e r and health p 2017 Insecticidal effects of plasma treated water **14** 1460

[14] Subramanian P G, Jain A, Shivapuji A M, Sundaresan N R, Dasappa S, Rao L J P P and Polymers 2020 Plasma-activated water from a dielectric barrier discharge plasma source for the selective treatment of cancer cells **17** 1900260

[15] Subramanian P G, Rao H, Shivapuji A M, Girard-Lauriault P-L and Rao L J J o A P 2021 Plasma-activated water from DBD as a source of nitrogen for agriculture: Specific energy and stability studies **129** 093303

[16] Rashid M, Rashid M, Reza M, Talukder M J P C and Processing P 2021 Combined Effects of Air Plasma Seed Treatment and Foliar Application of Plasma Activated Water on Enhanced Paddy Plant Growth and Yield **41** 1081-99

[17] Sajib S A, Billah M, Mahmud S, Miah M, Hossain F, Omar F B, Roy N C, Hoque K M F, Talukder M R, Kabir A H J P C and Processing P 2020 Plasma activated water: The next generation eco-friendly stimulant for enhancing plant seed germination, vigor and increased enzyme activity, a study on black gram (Vigna mungo L.) **40** 119-43

[18] Wartel M, Faubert F, Dirlau I, Rudz S, Pellerin N, Astanei D, Burlica R, Hnatiuc B and Pellerin S J J o A P 2021 Analysis of plasma activated water by gliding arc at atmospheric pressure: Effect of the chemical composition of water on the activation **129** 233301

[19] Kutasi K, Popović D, Krstulović N, Milošević S J P S S and Technology 2019 Tuning the composition of plasma-activated water by a surface-wave microwave discharge and a kHz plasma jet **28** 095010

[20] Apasheva L, Danyleiko Y, Egorov A, Korshunov A, Demin D, Sidorov V, Shilin L and Bogun I 2019 Activation of aqueous solutions by high-frequency glow discharge plasma in water vapor to stimulate growth and control diseases of agricultural plants. In: *IOP Conference Series: Earth and Environmental Science*: IOP Publishing) p 012027

[21] Van Gils C, Hofmann S, Boekema B, Brandenburg R and Bruggeman P J J o P D A P 2013 Mechanisms of bacterial inactivation in the liquid phase induced by a remote RF cold atmospheric pressure plasma jet **46** 175203

[22] Anaghizi S J, Talebizadeh P, Rahimzadeh H and Ghomi H J I t o p s 2015 The configuration effects of electrode on the performance of dielectric barrier discharge reactor for NO x removal **43** 1944-53

[23] Takaki K, Shimizu M, Mukaigawa S and Fujiwara T J I T o P S 2004 Effect of electrode shape in dielectric barrier discharge plasma reactor for NOx removal **32** 32-8

[24] Mei D and Tu X J J o C U 2017 Conversion of CO2 in a cylindrical dielectric barrier discharge reactor: Effects of plasma processing parameters and reactor design **19** 68-78

[25] Ozkan A, Dufour T, Bogaerts A, Reniers F J P S S and Technology 2016 How do the barrier thickness and dielectric material influence the filamentary mode and CO2 conversion in a flowing DBD? **25** 045016

[26] Nur M, Restiwijaya M, Muchlisin Z, Susan I, Arianto F and Widyanto S 2016 Power consumption analysis DBD plasma ozone generator. In: *Journal of Physics: Conference Series*: IOP Publishing) p 012101

[27] Wang X, Luo H, Liang Z, Mao T, Ma R J P S S and Technology 2006 Influence of wire mesh electrodes on dielectric barrier discharge **15** 845

[28] Chang M B and Wu S-J 1997 Experimental study on ozone synthesis via dielectric barrier discharges

[29] Subedi D, Tyata R, Shrestha R and Wong C 2014 An experimental study of atmospheric pressure dielectric barrier discharge (DBD) in argon. In: *AIP Conference Proceedings*: American Institute of Physics) pp 103-8

[30] Wang C and He X J A S S 2006 Effect of atmospheric pressure dielectric barrier discharge air plasma on electrode surface **253** 926-9

[31] Fang Z, Xie X, Li J, Yang H, Qiu Y and Kuffel E J J o P D A P 2009 Comparison of surface modification of polypropylene film by filamentary DBD at atmospheric pressure and homogeneous DBD at medium pressure in air **42** 085204

[32] Jahanmiri A, Rahimpour M, Shirazi M M, Hooshmand N and Taghvaei H J C e j 2012 Naphtha cracking through a pulsed DBD plasma reactor: Effect of applied voltage, pulse repetition frequency and electrode material **191** 416-25

[33] Qayyum A, Zeb S, Ali S, Waheed A, Zakaullah M J P c and processing p 2005 Optical emission spectroscopy of abnormal glow region in nitrogen plasma **25** 551-64

[34] Shemansky D, Broadfoot A J J o Q S and Transfer R 1971 Excitation of N2 and N+ 2 systems by electrons—I. Absolute transition probabilities **11** 1385-400

[35] Liu K, Ren W, Ran C, Zhou R, Tang W, Zhou R, Yang Z and Ostrikov K K J J o P D A P 2020 Long-lived species in plasma-activated water generated by an AC multi-needle-to-water discharge: effects of gas flow on chemical reactions **54** 065201

[36] Lim J S, Kim R H, Hong Y J, Lamichhane P, Adhikari B C, Choi J and Choi E H J R i P 2020 Interactions between atmospheric pressure plasma jet and deionized water surface **19** 103569

[37] Roy N C, Pattyn C, Remy A, Maira N, Reniers F J P P and Polymers 2021 NOx synthesis by atmospheric-pressure N2/O2 filamentary DBD plasma over water: Physicochemical mechanisms of plasma–liquid interactions **18** 2000087

[38] Sergeichev K F, Lukina N A, Sarimov R M, Smirnov I G, Simakin A V, Dorokhov A S and Gudkov S V J F i P 2021 Physicochemical Properties of Pure Water Treated by Pure Argon Plasma Jet Generated by Microwave Discharge in Opened Atmosphere 596

[39] Ghimire B, Szili E J, Patenall B L, Lamichhane P, Gaur N, Robson A J, Trivedi D, Thet N T, Jenkins A T A, Choi E H J P S S and Technology 2021 Enhancement of hydrogen peroxide production from an atmospheric pressure argon plasma jet and implications to the antibacterial activity of plasma activated water **30** 035009

[40] Moravej M, Yang X, Barankin M, Penelon J, Babayan S, Hicks R J P S S and Technology 2006 Properties of an atmospheric pressure radio-frequency argon and nitrogen plasma **15** 204

[41] Goldstein S, Lind J and Merényi G J C r 2005 Chemistry of peroxynitrites as compared to peroxynitrates **105** 2457-70